\documentclass[useAMS,usenatbib,usegraphicx,usedcolumn]{mn2e}

\newcommand{\CIV}{\mbox{C\,{\sc iv}}}
\newcommand{\CII}{\mbox{C\,{\sc ii}}}
\newcommand{\SiIV}{\mbox{Si\,{\sc iv}}}
\newcommand{\AlIII}{\mbox{Al\,{\sc iii}}}

\newcommand{\NV}{\mbox{N\,{\sc v}}}
\newcommand{\OVI}{\mbox{O\,{\sc vi}}}
\newcommand{\MgII}{\mbox{Mg\,{\sc ii}}}

\newcommand{\Ho}{\mbox{H$^0$}}
\newcommand{\Lya}{\mbox{Ly\,$\rm\alpha$}}
\newcommand{\Lyb}{\mbox{Ly\,$\rm\beta$}}
\newcommand{\cl}{{\sc CLOUDY}}
\newcommand{\kms}{km~s$^{-1}$}
\newcommand{\cms}{cm~s$^{-1}$}
\newcommand{\cmmt}{cm$^{-2}$}
\newcommand{\cmt}{cm$^{-3}$}
\newcommand{\Mi}{$M_{\it i}$}
\newcommand{\Nion}{\mbox{$N({\rm ion})$}}
\newcommand{\NHo}{\mbox{$N(\Ho)$}}
\newcommand{\NSiIV}{\mbox{$N(\SiIV)$}}

\newcommand{\nh}{\mbox{$n_{\rm H}$}}
\newcommand{\fnrepeat}[1]{$^{\ref{#1}}$}

\title[Metal enrichment by radiation pressure in AGN outflows]{Metal enrichment by radiation pressure in active galactic nucleus outflows -- theory and observations}
\author[A.~Baskin and A.~Laor]{Alexei Baskin\thanks{E-mail: alexei@physics.technion.ac.il (AB); laor@physics.technion.ac.il (AL)} and Ari Laor\footnotemark[1] \\ Physics Department, Technion -- Israel Institute of Technology, Haifa~32000, Israel}

\begin{document}
\date{}
\pagerange{\pageref{firstpage}--\pageref{lastpage}} \pubyear{2012}
\maketitle
\label{firstpage}

\begin{abstract}
Outflows from active galactic nuclei may be produced by absorption of continuum radiation by UV resonance lines of abundant metal ions, as observed in broad absorption line quasars (BALQs).  The radiation pressure exerted on the metal ions is coupled to the rest of the gas through Coulomb collisions of the metal ions. We calculate the photon density and gas density which allow decoupling of the metal ions from the rest of the gas. These conditions may lead to an outflow composed mostly of the metal ions. We derive a method to constrain the metals/H ratio of observed UV outflows, based on the \Lya\ and \SiIV~$\lambda\lambda$1394, 1403 absorption profiles. We apply this method to an SDSS sample of BALQs to derive a handful of candidate outflows with a higher than solar metal/H ratio. This mechanism can produce ultra fast UV outflows, if a shield of the continuum source with a strong absorption edge is present.
\end{abstract}

\begin{keywords}
galaxies: active -- quasars: absorption lines -- quasars: general.
\end{keywords}

\section{Introduction}
Broad absorption line quasars (BALQs) are a subclass of active galactic nuclei (AGNs) with spectra which exhibit broad, strong, and blue-shifted absorption features (\citealt{reichardetal03} and references there in). Approximately 15 per cent of quasars are classified as BALQs \citep{reichardetal03, kniggeetal08, sca09}. Although there are some differences in the continuum and emission lines between BALQs and non-BALQs, they appear to be drawn from the same parent population \citep{weymannetal91, reichardetal03}. The broad absorption lines (BALs) are the most prominent spectral feature indicating an outflow of material driven from and by the AGN. This outflow is a compelling candidate for the feedback mechanism of the AGN on its host. The outflow is commonly invoked as a growth and activity regulator (\citealt*{dimatteoetal05} and citations thereafter) and as a mechanism of metal enrichment of the ISM and IGM \citep{molletal07}.

How do the outflows form? Although much theoretical work has been done to model AGN outflows, a work which stems from studies of stellar winds (e.g., \citealt{lamcas99}), the driving physical mechanism of outflows remains uncertain. The first models involved only gas and radiation pressure as a driving force of the outflow (\citealt{vitshl88,arali94}; \citealt*{aravetal94}; \citealt{murrayetal95}; \citealt*{progaetal98}; \citealt{chenet01}). Later work introduced more detailed models of radiation pressure (\citealt*{progaetal99}; \citealt{chenet03a}) and magnetic fields \citep{dekbeg95}. The effect of radiation-driven wind on the emerging spectrum \citep{arabeg94} and galactic bulge \citep*{fabianetal06} has also been considered. More recently, with the advance in computational power, more detailed magnetohydrodynamical and magnetocentrifugal models of the outflow have been developed (\citealt{konkar94,proga00}; \citealt*{progaetal00}; \citealt{proga03, everett05, kurpro08}). Currently, the models produce almost no predictions which can help discriminate between them using observations (see a review by \citealt*{crenshawetal03}).

In current models the outflow is treated as a single-component fluid, and the
radiation pressure is parametrized by a ``force multiplier'', which encapsulates the effects of radiation pressure due to Thomson scattering and line and dust absorption. What effect does the radiation pressure have on the different constituents of the fluid? As was first pointed out by \citet{sp92}, for radiation-driven stellar winds, a single-component fluid treatment overlooks the possibility of metal ions decoupling from the mostly-H outflow. \citet{sp92} have shown that metal ions can run away (i.e., decouple) from the proton-electron fluid, if the coupling between ions and the fluid is small enough, compared to the photon flux absorbed by the ions. Photon absorption by a metal ion results in a net gain of momentum i.e., ion acceleration (assuming illumination by an anisotropic source), while the acceleration of H is inefficient as H is mostly ionized. The ion is decelerated by Coulomb scatterings by the ambient protons and electrons. The deceleration is velocity dependent and is maximal when the ion has a velocity similar to the thermal velocity of each of the constituents of the proton-electron fluid. The ion can run away from the fluid, if the rate of photon absorption (i.e., flux density) is large enough, so that the accumulated gain in velocity of the ion surpasses the fluid thermal velocity before the ion is significantly deflected by Coulomb scatterings i.e., if the acceleration due to the radiation pressure is larger than the deceleration due to the Coulomb force. The metal decoupling prevents the proton-electron fluid from accelerating further and the decoupled metals will produce a fast wind through the proton-electron fluid.

Metal ion decoupling from the mostly-H gas may be relevant in AGN outflows due to two reasons. First, gas in close vicinity of the active nucleus is exposed to a large flux of radiation; second, strong ion absorption is observed. Our goal is to explore whether the occurrence of metal decoupling is a plausible process in AGN. Self-consistent models of an outflow and the structure of the ionized gas region, which incorporate the process of metal ion decoupling, are beyond the scope of this study.

How can an occurrence of the metal decoupling be observationally verified? The metal decoupling will mostly affect the absolute metallicity of a wind i.e., metal to H abundance ratio. The decoupling effect on the relative i.e., metal to metal abundance ratio, might be negligible. Since the wind can be composed from several types of decoupled metals, the relative abundance ratio of the wind might be similar to that of the wind origin, where the metals are mixed with H. Thus, indirect (i.e., relative) outflow metallicity measurements cannot be used to test the proposed metal enrichment scenario. Direct (i.e., absolute metals to H) metallicity measurements are needed to test the validity of the enrichment scenario in AGN.

There is accumulating evidence of metal abundances higher than solar for the broad line region (BLR; \citealt{hamann97}, \citealt{hamfer99}). There is also evidence for higher than solar metal abundances in AGN outflows from analysis of narrow absorption lines (\citealt{aravetal07, wuetal10, hamannetal11}) and broad absorption lines (\citealt{hamannetal97}). However, for some objects the results are inconclusive (\citealt{hamann98}) or consistent with solar abundances (\citealt{aravetal01}). The main difficulty in estimating the element abundance arises from a large uncertainty in the measurement of ionic column densities when the covering factor is velocity dependent (\citealt{hamann98, aravetal99}). Note however that \citet{cottisetal10} disfavour radiation-driven outflow altogether, due to the absence of an excess of objects displaying line-locking between \Lya\ and \NV. A metallicity of a few times solar can be explained by stellar enrichment due to extended star formation (e.g., \citealt{hamannetal02}), as suggested in high-redshift quasars ($z\geq3.5$, \citealt{dietrichetal03}). However, if a metal ions runaway occurs, the metal/H column ratio can be larger by orders of magnitude, compared to the solar abundance ratio.

Below we derive the conditions for an ion runaway in gas in the vicinity of an active nucleus. We also present a method to set a robust and direct lower limit on the metal/H ratio in outflows based on UV absorption lines. The method is then implemented on a sample of AGN from the Sloan Digital Sky Survey (SDSS; \citealt{york00}) BALQs catalog \citep{sca09}. The ion runaway model is outlined in Sec.~2. The direct metallicity estimation method and its implementation are described in Secs.~3 and 4, respectively. We discuss the results in Sec.~5. Our conclusions are summarized in Sec.~6.

\section{METAL IONS RUNAWAY}\label{sec:metal}
We begin by describing the conditions for the separation of metal ion from an ionized H gas (i.e., runaway). First, we estimate the photon absorption rate by the metal ion, and the mean momentum gain by the absorption of each photon. A product of the two quantities is equal to the accelerating force on the ion due to radiation pressure. Second, this radiative accelerating force is compared to the decelerating force that is caused by Coulomb scattering of the metal ion by the H gas (i.e., protons and electrons). Finally, we investigate the conditions (photon flux density, H number density and temperature) that lead to an accelerating force larger than the decelerating one. We first present a qualitative order-of-magnitude estimates, and then a more exact and formal derivation of the separation mechanism.

\subsection{Qualitative estimates}
An estimate of the ionization parameter $U$ for a metal ion runaway in AGN can be roughly estimated as follows. The mean momentum gain of an ion due to absorption of a photon is
\begin{equation}
\bar{p}=\frac{h}{\lambda_{\rm ion}},\label{eq:p_gain}
\end{equation}
where $h$ is Plank's constant and $\lambda_{\rm ion}$ is the wavelength of the absorption transition. The mean increment of the ion velocity due to photon absorption is $\delta v=\bar{p}/m_{\rm ion}$, where $m_{\rm ion}$ is the ion mass. A typical value for an UV absorption line, e.g.~\CIV, is $\delta v\approx 20$~\cms. The photon absorption rate is 
\begin{equation}
r_\gamma=B_{12}\bar{J},\label{eq:rate_gam}
\end{equation} 
where $B_{12}$ is the Einstein absorption coefficient\footnote{The Einstein absorption coefficient is given by $B_{12}=4\pi^2 e^2 \lambda_{\rm ion}f_{12}/h m_e c^2$, where $e$ and $m_e$ are the electron charge and mass, $f_{12}$ is the transition oscillator strength and $c$ is the speed of light. We adopt the values of $f_{12}$ and $\lambda_{\rm ion}$ from \citet{mor91}.}, and $\bar{J}$ is the mean intensity $J_\nu$ averaged over the metal ion absorption line profile i.e., $\bar{J}=\int^{\infty}_{0} J_\nu\phi(\nu)d\nu$, where $\phi(\nu)$ is the line profile function (e.g., \citealt{ryblih04}). The value of $B_{12}$ for typical resonance UV lines (e.g., \CIV) is $\sim7\times10^{9}$~cm$^2$~erg$^{-1}$~s$^{-1}$. We begin by estimating $r_\gamma$ in the BLR in AGN. We assume a bolometric luminosity of $L_{\rm bol}=10^{46}$~erg~s$^{-1}$ and $\nu L_\nu=2.5\times 10^{45}$~erg~s$^{-1}$ at $\lambda=1500$~\AA, which yields $L_\nu(1500~\mbox{\AA})=1.3\times10^{30}$~erg~s$^{-1}$~Hz$^{-1}$. The implied BLR distance from the central illuminating source is $R_{\rm BLR}=0.1$~pc, and the resulting mean intensity is $\bar{J}\approx J_\nu(1500~\mbox{\AA})\approx10^{-7}$~erg~s$^{-1}$~cm$^{-2}$~ster$^{-1}$~Hz$^{-1}$. Thus, the photon absorption rate is $r_\gamma\sim7\times10^{9}\cdot1\times10^{-7}\approx10^3$~s$^{-1}$.

The decelerating force acting on the metal ion is due to Coulomb collisions with the ionized H gas constituents i.e., protons and electrons. A simplified condition for a runaway is that the metal ion gains a velocity larger than the thermal velocity of the ions $v_{\rm th,ion}$ between consecutive collisions. Mathematically this condition can be stated as 
\begin{equation}
\delta v\, r_\gamma/r_{\rm col}>v_{\rm th,ion}, \label{eq:simpl_cond}
\end{equation} 
where $r_{\rm col}$ is the collision rate. Noting that the main contribution to the decelerating force at low ion velocities originates from collisions with protons (see Sec.~\ref{sec:form_desc}), we consider collisions only with protons for the remainder of this qualitative description. The collision rate is
\begin{equation}
r_{\rm col}=n_{\rm p} v_{\rm th,p} \sigma_{\rm col},\label{eq:r_col}
\end{equation}  
where $n_{\rm p}$ is the number density of the protons, $v_{\rm th,p}$ is the proton thermal velocity and $\sigma_{\rm col}$ is the cross-section for a proton-ion collision. The collision cross-section can be estimated by evaluating the impact parameter $b$ at which the metal ion is scattered by 90~deg. The 90~deg scattering occurs when the Coulomb potential energy and the kinetic energy of the ion ($E_{\rm kin}$) are equal i.e., $Z_{\rm ion}e^2/b\sim E_{\rm kin}$, where $e$ is the electron charge and $Z_{\rm ion}$ is the ionization state of the ion ($\sim3$). We take $E_{\rm kin}$ to be the thermal energy i.e., $E_{\rm kin}\approx kT\sim~10^{-12}$~erg for a typical BLR temperature of $T=10^4$~K, where $k$ is the Boltzmann constant. This yields $b\approx5\times10^{-7}$~cm and $\sigma_{\rm col}=\pi b^2\approx8\times10^{-13}$~cm$^2$. The ion and protons thermal velocities for $T=10^4$~K are $v_{\rm th,ion}\approx4\times10^5$~\cms\ and $v_{\rm th,p}\approx1\times10^6$~\cms; and from Eq.~\ref{eq:simpl_cond}, $r_{\rm col}<20\times10^3/4\times10^5\sim5\times10^{-2}$~s$^{-1}$. Now we can find the condition on the number density from Eq.~\ref{eq:r_col}, $n_{\rm p}<5\times10^{-2}/(1\times10^6\cdot 8\times10^{-13})\sim10^5$~\cmt\ in order to get a runaway acceleration. 

Adopting the accelerating and decelerating forces estimated above, one can evaluate $U_{\rm r}\equiv n_\gamma/n_{\rm p}$ that produces a runaway, where $n_\gamma$ is the ionizing flux density. Since $n_{\rm p}=10^{10}$~\cmt\ is the typical $U=0.1$ BLR density, one needs $U_{\rm r}=0.1\times(10^{10}/10^5)\approx10^4$ to have a runaway. The value of $U_{\rm r}$ is
independent of $R$, since $r_\gamma\propto R^{-2}$, and thus $n_{\rm p}\propto R^{-2}$, and also $n_\gamma\propto R^{-2}$.
At this large value of $U_{\rm r}$, photoionized gas is fully ionized, and the relevant UV absorbing metal ions are not present.  A runaway thus requires a shield between the illuminating source and the gas. The shield has to filter the high energy photons that can overionize the metals (e.g., $>4.7$~Ryd for C-outflow).\footnote{1~Ryd = 13.6~eV} Note that such a shield appears to be required to explain radiation-driven BALQ-outflows (e.g., \citealt{murrayetal95}).

\subsection{Quantitative estimates}\label{sec:form_desc}
We describe a microscopic model in which an ion (e.g., C$^{3+}$) is embedded in a fully ionized gas mostly made of ionized H, illuminated by a continuum in the radial direction. In addition, we make the following simplifying assumptions:
\begin{enumerate}
\item The ion is assumed to have only one transition i.e., only one absorption line at the laboratory wavelength $\lambda_{\rm ion}$.
\item All calculations are conducted in the non-relativistic limit.
\item The gas is in a free-fall circular motion, held by gravity and radiation pressure.
\end{enumerate}   
The mean momentum transformed from a photon to an ion in a process of absorption followed by reemission is given by Eq.~\ref{eq:p_gain}. The photon absorption rate (Eq.~\ref{eq:rate_gam}) expressed in terms of oscillator-strength $f_{12}$ and flux density $f_\nu$ (note that $J_\nu=f_\nu/4\pi$) is
\begin{equation}
r_\gamma=\frac{\pi e^2}{m_e chc/\lambda_{\rm ion}}f_{12}f_\nu(\lambda_{\rm ion}).\label{eq:rate_gam2}
\end{equation}
The acceleration due to radiation pressure is
\begin{equation}
a_{\rm rad}=\bar{p}/m_{\rm ion}\cdot r_\gamma=\frac{\pi e^2}{m_e c^2}\frac{f_{12}}{m_{\rm ion}}f_\nu(\lambda_{\rm ion}).\label{eq:a_rad}
\end{equation}
The main relevant UV transitions, \CIV, \SiIV\ and \NV, have similar atomic parameters ($f_{12}/m_{\rm ion}\sim10^{22}$~gr$^{-1}$) yielding
\begin{equation}
a_{\rm rad}\sim10^{10}f_\nu(\lambda_{\rm ion})\,\mbox{[cm s$^{-2}$]}.
\end{equation}

We adopt the method described by \citet{sp92} for multicomponent fluid winds driven by radiation of hot stars.\footnote{The method can be derived from the ``classical'' description of Coulomb friction \citep{spitz62} by adopting a reduced mass instead of the field mass in the equation for velocity normalization (denoted by $l_{\rm f}$ in Spitzer). This adaptation is needed because the derivation in Spitzer is made for an electron and not for an ion particle. The reduced mass with a proton is $\approx m_e$ for the former, but it is $\approx m_p$ for the latter, where $m_p$ is the proton mass.} Protons and electrons induce a deceleration force on the test-particle due to Coulomb friction \citep[e.g.,][]{spitz62}. The total deceleration is
\begin{equation}
a_{\rm fric}=\sum_{i={\rm p, e}}n_i\frac{4\pi e^4 Z_{\rm ion}^2}{k T m_{\rm ion}}\ln\Lambda G(x_i),\label{eq:fric}
\end{equation}
where $n_i$ is the number density of constituent $i$ (note that $n_{\rm e}\approx n_{\rm p}\approx n_{\rm H}$, where the latter is the H number density), $Z_{\rm ion}$ is the ionization state of the test particle, $\ln\Lambda$ is the Coulomb logarithm, $G(y)=[\Phi(y)-y \Phi^\prime (y)]/2y^2$ is the Chandrasekhar function with $\Phi(y)$ being the error function and
\begin{equation}
x_i=\sqrt{\frac{m_{\rm ion}m_i}{m_{\rm ion}+m_i}}\frac{w}{\sqrt{2kT}},
\end{equation}
where $w$ is the relative velocity between the test particle and the fluid \citep[e.g.,][]{spitz62}. For the Coulomb logarithm we adopt the equation from \citet{spitz62},
\begin{equation}
\ln\Lambda=\ln\frac{3}{2Z_{\rm ion}e^3}\sqrt{\frac{k^3 T^3}{\pi n_{\rm H}}}.\label{eq:ln_L}
\end{equation}
Several authors use the total number density of free particles instead of $n_{\rm H}$ \citep[e.g.,][]{owo02}. This has a minor effect on $\ln\Lambda$, lowering it by few per cents. 

\begin{figure}
\includegraphics[width=84mm]{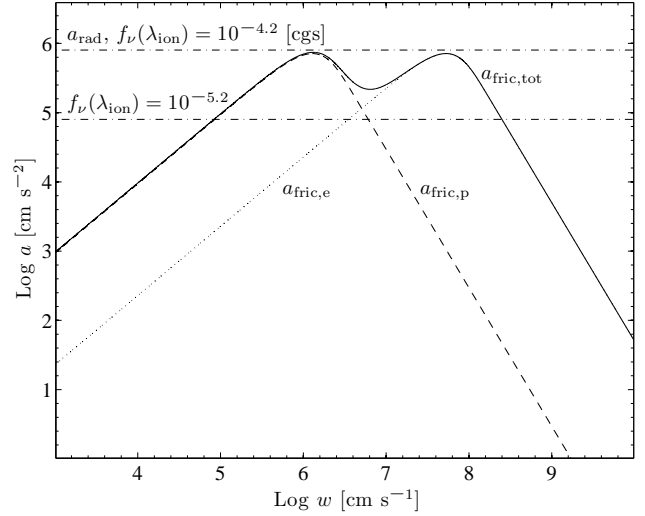}
\caption{The Coulomb friction deceleration as a function of relative velocity between a test-particle and a fluid. The deceleration due to protons (dashed line), electrons (dotted line), and the total deceleration (solid line) are presented for $n_{\rm H}=10^6$~\cmt, $T=10^4$~K and a C$^{3+}$-like test particle (Eq.~\ref{eq:fric}). The radiative acceleration of a C$^{3+}$ particle absorbing at \CIV~$\lambda$1549 transition (dot-dashed line) is also plotted for two different $f_\nu(\mbox{1549~\AA})$. Note that the maximal deceleration of each fluid constituent is achieved when $w\approx v_{\rm th}$, where $v_{\rm th}=\sqrt{2kT/m_{\rm p,e}}$ is the thermal velocity of the constituent.  An ion will run away when $a_{\rm rad}>a_{\rm fric}$.}\label{fig:dyn_fric}
\end{figure}

Figure~\ref{fig:dyn_fric} presents the Coulomb friction deceleration as a function of the relative velocity $w$, while assuming $n_{\rm H}=10^6$~\cmt, $T=10^4$~K and a C$^{3+}$-like test-particle ion. We plot both the individual decelerations due to protons and electrons, and the total friction deceleration. Protons contribute to $a_{\rm fric}$ mostly for $w\la 10^7$~\cms, while electrons contribute above those velocities. 

The maximal frictional deceleration ($a_{\rm fric}^{\rm max}$) for each constituent is reached when $w\approx v_{\rm th}$, where $v_{\rm th}$ is the thermal velocity of the constituent. This is because $G(x_i)$ peaks at $x_i\approx1$ [$G(1)=0.214$ (e.g., \citealt{kul05})], and $m_{\rm ion}\gg m_{\rm e}, m_{\rm p}$ which yields $x_i\approx w/\sqrt{2kT/m_i}=w/v_{\rm th, i}$. Once a test particle reaches $w>v_{\rm th, e}$, the total frictional decelerating force on it diminishes and it runs away (if the accelerating force remains active). Since the dependence of $\ln\Lambda$ on $T$ and $n_{\rm H}$ is logarithmic (Eq.~\ref{eq:ln_L}), and $T$ and $n_{\rm H}\gg1$ for an AGN environment, $\ln\Lambda\approx\mbox{const}$. This approximation simplifies the $a_{\rm fric}^{\rm max}$ dependence on $T$ and $n_{\rm H}$ to $a_{\rm fric}^{\rm max}\propto n_{\rm H} T^{-1}$. The $a_{\rm fric}^{\rm max}$ can be estimated to a good approximation by
\begin{equation}
\log a_{\rm fric}^{\rm max} \approx (3.8^{+0.5}_{-0.8}) + \log\frac{\nh}{T}\mbox{ [cm s$^{-2}$]},
\end{equation}
where the range on the constant is the maximal variation of Eq.~\ref{eq:fric} [taking $G=G(1)$] for the C$^{3+}$, N$^{4+}$ and Si$^{3+}$ ions and $\ln\Lambda$, when \nh\ is varied in the $10^3$--$10^{12}$~\cmt\ range and $T$ in the $10^3$--$10^6$~K range. Note that for $T=10^6$~K $v_{{\rm th},e}\approx0.02c$ and raising $T$ to much higher values will break the non-relativistic approximation.\footnote{Note that for a very large temperature ($T\ga10^8$~K) one needs to use the de Broglie wavelength $\lambda_{\rm dB}$ as the distance of minimal approach in the calculation of $\Lambda$ (e.g., \citealt{kul05}). Even for $T\ga10^5$~K one needs to use $\lambda_{\rm dB}$ for interactions with electrons. This introduces a small {\it negative} correction to $\ln\Lambda$, which causes $a_{\rm fric,e}^{\rm max}<a_{\rm fric,p}^{\rm max}$. Thus, disregarding this effect does not affect the results of our study.}

An ion will run away when $a_{\rm rad}>a_{\rm fric}^{\rm max}$, i.e. when
\begin{equation}
\log f_\nu(\lambda_{\rm ion})-\log\nh+ \log T>-6.2.\label{eq:runaway}
\end{equation}
Assuming a spectral energy distribution (SED), we can convert Eq.~\ref{eq:runaway} to a condition on $U$. An SED with $f_\nu\propto\nu^{-1}$ up to 1~Ryd, and a steeper slope beyond that, gives $n_\gamma\approx\nu f_\nu/h\nu c\approx5 \times10^{15}f_\nu(\mbox{1~Ryd})\approx3\times10^{15}f_\nu(\lambda_{\rm ion})$. Inserting in Eq.~\ref{eq:runaway} yields
\begin{equation}
\log n_\gamma -15.5-\log\nh+\log T>-6.2\label{eq:runaway_U}
\end{equation}
or simply, $\log U>5.3$ for $\log T=4$.

The model described above and the derived conditions are only estimates of the conditions needed to produce a runaway of metals. One needs to sum up over all resonance lines. Also, the
ionization state of the metal ions changes with time, and the relevant resonance lines change.
For example, a C atom is found $\sim 60$ per cent of the time in the C$^{3+}$ ionization state, when the latter is the dominant state (e.g., Fig.~\ref{fig:ion_struc1}, top panel). Note that the electric field produced by the separation of a metal ion causes the ion to ``carry'' with it the proper number of electrons, and the gas remains globally neutral.

The difference in ionization fractions of metals compared to H is the main reason why the radiative acceleration is more efficient for metals than for H. When the gas is ionized so that there is a significant fraction of elements that are ionized three and above times (C$^{3+}$, Si$^{3+}$, N$^{4+}$, etc.), H atoms are mostly ionized and the fraction of H$^0$ is $\sim 10^{-5}U^{-1}$ (e.g., \citealt{ferland99}; see also Fig.~\ref{fig:ion_struc1}, top panel). The $r_\gamma$ for H$^0$ and metal ions is comparable, but the metal ions spend most of the time at the dominant ionization state, while H spends only a fraction of $10^{-5} U^{-1}$ of the time as H$^0$, and being accelerated. Therefore, the relative $a_{\rm rad}$ of H is reduced by a factor of $10^{-5} U^{-1}$ and is negligible.

\section{Estimating H/metal column ratio in outflows} \label{sec:method}
The straightforward method to directly estimate the absolute metallicity of an outflow is to measure the ionic column density ratio of metals to H. The difficulty to implement this method for quasars is that usually the H column density, derived based on \Lya\ absorption, is poorly constrained. Most spectra are obtained from ground based observation of $z\ga3$ quasars, which are subject to significant intervening \Lya\ absorption systems. This foreground absorption makes it difficult to determine the unabsorbed emission level. In addition, the potentially absorbed blue wing of \Lya\ cannot be estimated based on the unabsorbed red wing, as the red wing is blended with the \NV\ emission line and is also often partly absorbed by the \NV\ absorber. Thus, we choose to constrain only the minimal allowed \NHo\ in a statistical sense by utilizing composite spectra of non-BALQs (see Sec.~\ref{sec:data}).

The structure of this section is as follows. First, we search for the best metal absorption line that can be used to place a direct lower limit on the gas absolute metallicity (Sec.~\ref{sec:which_line}). Then, we adopt photoionization models to set a lower limit on the necessary \Lya\ absorption from that gas (Sec.~\ref{sec:photo_model}). Finally, we present the procedure for estimating a lower limit on the absolute metallicity i.e., the H/metal abundance ratio in the outflowing gas (Sec.~\ref{sec:prescr}).

\subsection{Which absorption line to use?}\label{sec:which_line}
To directly constrain the H abundance associated with a metal absorber, we need a UV metal line which is expected to be associated with a large \NHo\ column. Three criteria are imposed to find the best metal absorption line.
\begin{enumerate}
\item The line can be observed with \Lya\ at the same wavelength coverage of a typical instrument (e.g., the SDSS covers the 3800--9200~\AA\ range, and requires $z\ga2.7$ to cover the \Lya--\CIV~$\lambda1549$ wavelength region). This criterion disfavors \MgII~$\lambda$2799 which is found too far to the red relative to \Lya.
\item The line should be prominent and not blended with absorption lines of other metals. This criterion disfavors the remaining low ionization UV lines. Although C$^{+}$ (\CII~$\lambda$1335) and Al$^{++}$ (\AlIII~$\lambda$1857) produce larger column density ratios \NHo/\Nion\ than the higher ionization ions, their absorption profile is generally shallower, and the implied minimal \Lya\ absorption can be harder to detect. These lines are therefore excluded.
\item The photoionization models should imply the largest minimal \Lya\ optical depth $\tau$ for a given $\tau$ of metal line. Since $\tau\propto f_{12}\lambda_{\rm ion}\times N({\rm ion})$, this is equivalent to requiring a minimal $\NHo/\Nion\times1/f_{12}\lambda_{\rm ion}$ as high as possible. The value of \NHo/\Nion\ is calculated by photoionization models, and $f_{12}\lambda_{\rm ion}$ is set by the atomic physics. This criterion favours \SiIV\ (see Sec.~\ref{sec:photo_model_results}), which is observed to be as prominent as \CIV\ in absorption for the objects studied here.
\end{enumerate}
Thus, the above criteria favour by elimination the \SiIV\ absorption line. 

\subsection{The photoionization models}\label{sec:photo_model}
What is the minimal possible \NHo\ for a given \Nion? We estimate this quantity  using the photoionization code \cl\ for a very wide range of model parameters. Calculations are performed with version 08.01 of \cl\ \citep{fer98}. A plane parallel geometry is assumed. The gas is assumed to have a solar composition. The ionization parameter $U$ and the total H column density $\Sigma$ are varied by 0.5~dex in the following range: $-3\leq\log U\leq 2$ and $15\leq\log\Sigma\leq24$. The H density $n_{\rm p}$ is varied by 1~dex in the range $5\leq\log n_{\rm p}\leq12$. It should be noted that the assumed $n_{\rm p}$ has a small effect on the deduced minimal \NHo/\Nion. Models with $3\le\log U\le 5$, $\log \Sigma=25-26$ and $\log n_{\rm p}=8$ are also calculated to find an absolute minimum on \NHo/\Nion\ (see below). Note however that for these parameters the gas is optically thick to electron scattering ($\tau_{\rm e.sc.}>1$), and \cl\ results may be incorrect, since the code is not designed to simulate regimes with $\tau_{\rm e.sc.}$ well above 1.

\subsubsection{The spectral energy distribution}
The minimal possible value for \NHo/\Nion\ is mostly set by the assumed SED of the illuminating source. The value of \NHo/\Nion\ at a given $U$ and $n_{\rm p}$ is set by the number of photons able to ionize \Ho\ relative to number photons which produce the ion without ionizing it to the next ionization state. Since \Ho\ is mostly ionized by photons with energies at $\sim$1 -- 2~Ryd, a steeper SED in the 1 -- 5~Ryd range implies more \Ho\ ionizing photons for a given ionic column of Si$^{+3}$ and C$^{+3}$ (for a fixed $n_{\rm p}$). Thus, a steeper slope produces a lower \NHo/\Nion. 

We adopt a broken power-law SED for the incident continuum. We use the steepest SED consistent with observations to produce the minimal \NHo/\Nion. The adopted spectral slopes $\alpha$ ($f_\nu\propto\nu^{-\alpha}$) are described below. In the range 3000 -- 10,000~\AA, we use the slope measured by \citet{elvisetal94} (a cutoff is assumed above 1~$\mu$m). In the UV range (1000 -- 3000~\AA), we use the slope which is measured using the flux of the composite BALQ spectrum (see Sec.~\ref{sec:balq_data_set}) between $\sim1300$ and 2050~\AA. The measured $\alpha=1.8$ is extremely red compared to the typical AGN $\alpha=0.5$ in this wavelength range, but since the \NHo/$N(\mbox{Si$^{3+}$})$ depends on energies larger than 1~Ryd (i.e., $\lambda<912$~\AA), the results of this study are not sensitive to the precise value of $\alpha$ below 1~Ryd. In the 500 -- 1200~\AA\ range, \citet{tel02} find $\alpha$ of 2.0 and 1.6 for radio-loud (RL) and radio-quite (RQ) quasars, respectively. Although most BALQs are RQ \citep{stockeetal92,beckeretal10}, we adopt the steeper slope of RL quasars for $\lambda=500-1000$~\AA, because we study carefully selected objects which may happen to be RL. There are no direct measurements of $\alpha$ in the $\sim60-500$~\AA\ range, and we extend the \citet{tel02} slope down to 12~\AA\ (1~keV). Note that $\alpha=2$ is steeper than the slope reported by \citet{laoretal97} for the soft X-ray range (0.2 -- 1~keV; $\alpha=1.2$ and 1.7 for RL and RQ quasars, respectively), but this energy range has little effect on the results of our study. In the 2 -- 10~keV range, \citet*{brandtetal97} measure $\alpha=1$, and \citet{reetur00} report $\alpha=0.7$ and 0.9 for RL and RQ quasars, respectively. We adopt $\alpha=1$ for the whole hard X-ray range (1 -- 100~keV), and assume a cutoff above 100~keV \citep{molinaetal09}. The overall adopted SED has $\alpha_{\rm ox}=1.9$ between optical (3000~\AA) and X-ray (2~keV), which is close to the typical BALQ $\alpha_{\rm ox}=2$ \citep*{brandtetal00}. Table~\ref{tab:SED} summarizes the adopted slopes and the respective energy and wavelength ranges.

\begin{table}
\begin{minipage}{84mm}
\caption{The adopted spectral energy distribution.\fnrepeat{fn1:enrg}}\label{tab:SED}
\begin{tabular}{@{}*{3}{c}p{2.7cm}@{}}
\hline
Energy range & $\lambda$ range\fnrepeat{fn1:wave}\\ (eV) & (\AA) & $\alpha$\fnrepeat{fn1:a}& References \\
\hline
1.2 -- 4.1 & 3000 -- 10,000 & 0.5 & \citet{elvisetal94}\\
4.1 -- 12.4 & 1000 -- 3000  & 1.8 & this paper\fnrepeat{fn1:2}\\
12.4 -- 24.8 & 500 -- 1000 & 2.0 & \citet{tel02}\\
24.8 -- 1000 & 12 -- 500 & 2.0 & ---\fnrepeat{fn1:4}\\
$(1-100)\times10^3$& 0.1 -- 12 & 1.0 & \citet{brandtetal97}; \citet{reetur00}\fnrepeat{fn1:6}\\ 
\hline 
\end{tabular}
\footnotetext[1]{A cutoff is assumed below 1~$\mu$m and above 100~keV. The overall SED has a slope of $\alpha_{\rm ox}=1.9$ between 3000~\AA\ and 2~keV.\label{fn1:enrg}}
\footnotetext[2]{The corresponding wavelength range for the listed energy range.\label{fn1:wave}}
\footnotetext[3]{$f_\nu\propto\nu^{-\alpha}$.\label{fn1:a}}
\footnotetext[4]{The slope is the steepest one consistent with the composite BALQ spectrum between $\sim1300$ and 2050~\AA\ (see text). This slope is then extrapolated to 1000 and 3000~\AA.\label{fn1:2}}
\footnotetext[5]{Extrapolation of the \citet{tel02} slope.\label{fn1:4}}
\footnotetext[6]{Note that \citet{brandtetal97} and \citet{reetur00} measurements were conducted in the 2 -- 10~keV range.\label{fn1:6}}
\end{minipage}
\end{table}

\subsubsection{The model results}\label{sec:photo_model_results}
\begin{figure}
\includegraphics[width=84mm]{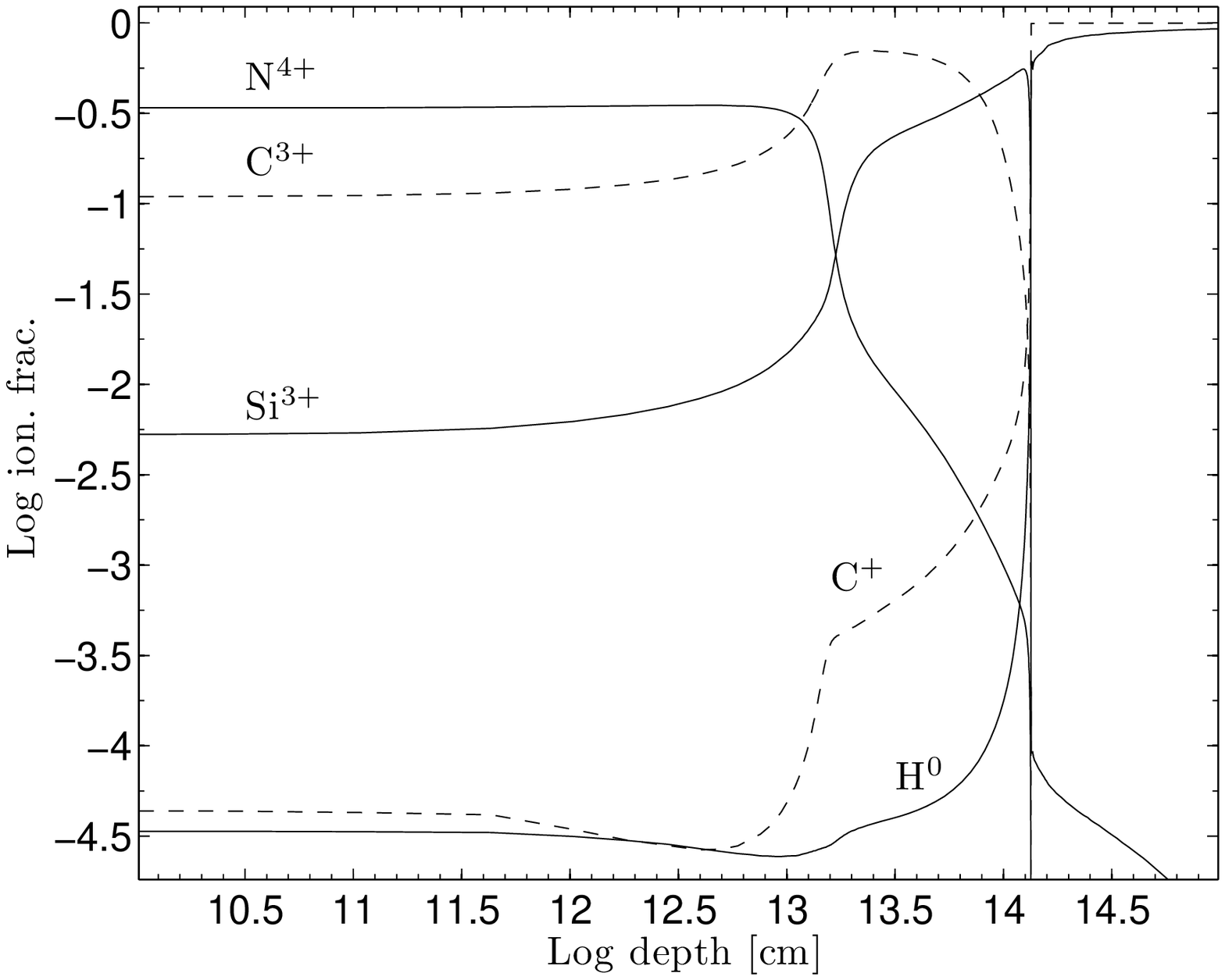}
\includegraphics[width=84mm]{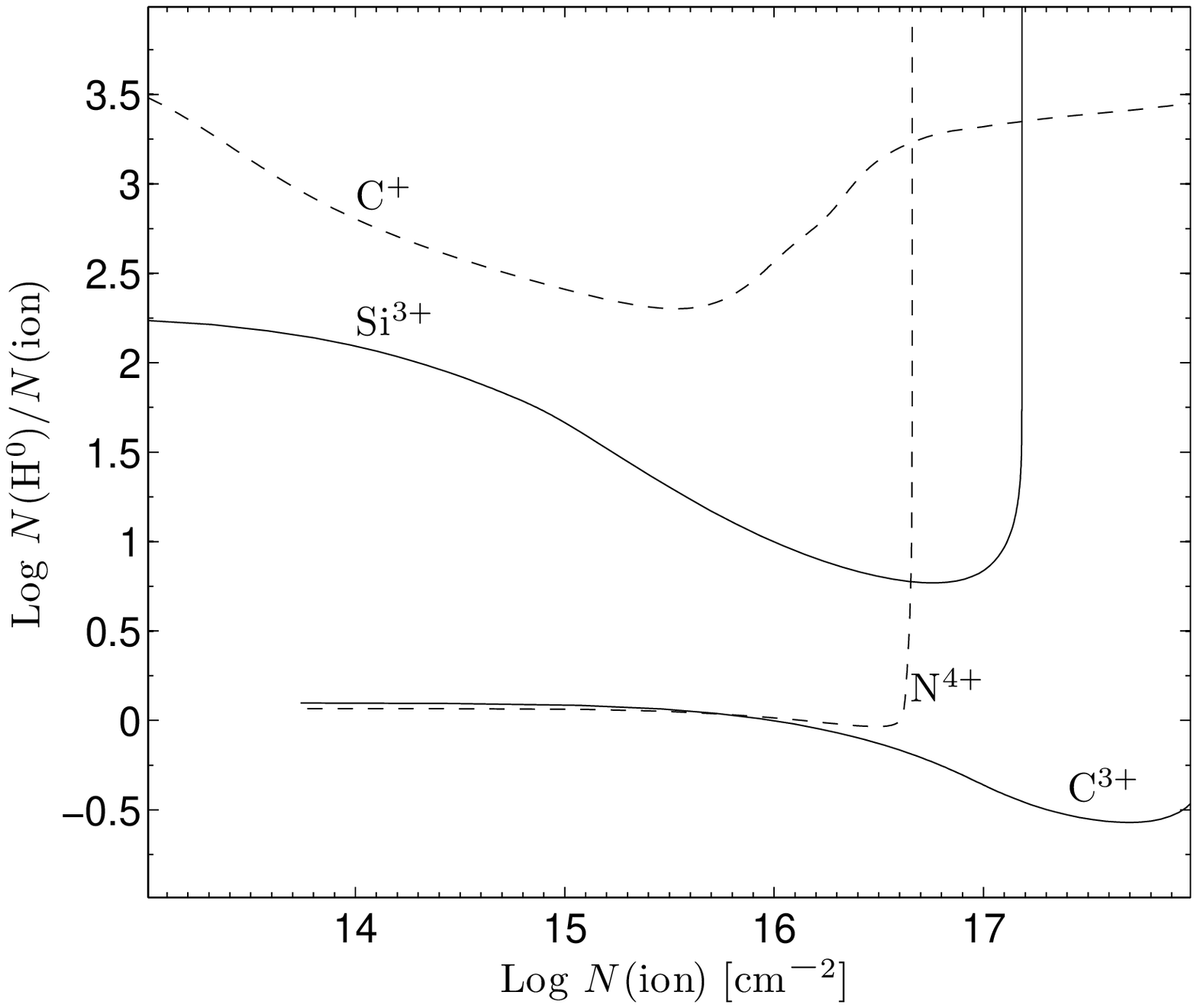}
\caption{The slab ionization structure (top) and the ratio of column of \Ho\ to the column of different ions versus the ion column (bottom). We assume $\log U=-1$ and $n_{\rm p}=10^8$~\cmt. The assumed abundances relative to H of C, N and Si are $-3.61$, $-4.07$ and $-4.46$ in log-scale (\cl\ default solar abundances). Although C$^+$ can place a higher constraint on the allowed $N(\Ho)$ than Si$^{3+}$, it is not used in this study, because the observed \CII\ absorption is weak.}\label{fig:ion_struc1}
\end{figure}

Figure~\ref{fig:ion_struc1} presents the ionization structure of the slab (top panel), and the integrated \NHo/\Nion\ as a function of integrated \Nion\ (bottom panel; this is equivalent to assuming different $\Sigma$), assuming $\log U = -1$ and $n_{\rm p}=10^8$~\cmt. Although the Si$^{3+}$ ionization fraction is larger by $\sim3.5$~dex than \Ho\ in the region where the Si$^{3+}$ has a significant fraction and column (top panel), the integrated \NHo/$N(\mbox{Si$^{3+}$})\sim10$ (bottom panel) because H is $\sim10^{4.5}$-time more abundant than Si (assuming $Z_{\sun}$). Note that C$^{3+}$ and N$^{4+}$ place a lower constraint on \NHo\ than Si$^{3+}$. A higher constraint is placed by C$^+$. But, the observed absorption profile of \CII~$\lambda$1335 is shallower than \SiIV~$\lambda$1397 (see Sec.~\ref{sec:data}) and yields a lower overall constraint on \Lya\ absorption. Thus, we chose Si$^{3+}$ as the optimal ion to constrain \NHo. 

\begin{figure}
\includegraphics[width=84mm]{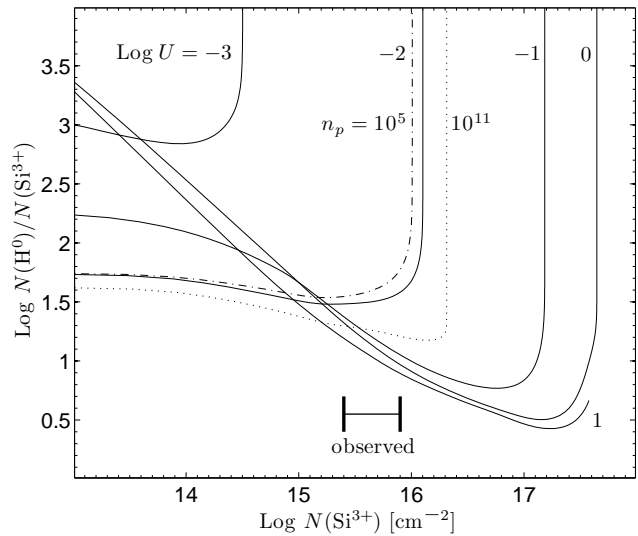}
\caption{The column $N(\Ho)$ relative to the column $N(\mbox{Si$^{3+}$})$ versus the latter for different ionization models. The assumed ionization parameters are indicated. All models are calculated with $n_{\rm p}=10^8$~\cmt\ (solid lines). Two additional models with $n_{\rm p}=10^5$ and $10^{11}$~\cmt\ are plotted for $\log U=-2$ (dot-dashed and dotted lines, respectively). Note the similarity between the three curves. The range of measured $N(\mbox{Si$^{3+}$})$ in this study for a single absorption system is marked by two thick vertical bars. Note that in this range $N(\Ho)$/$N(\mbox{Si$^{3+}$})\ga 8$ for all possible models.}\label{fig:ion_struc2}
\end{figure}

Figure~\ref{fig:ion_struc2} presents the dependence of \NHo/$N(\mbox{Si$^{3+}$})$ on $N(\mbox{Si$^{3+}$})$. The dependence is plotted for different values of $U$ assuming $n_{\rm p}=10^8$~\cmt. For $\log U = -2$ we also plot the dependence for $n_{\rm p}=10^5$ and $10^{11}$~\cmt, which produces similar curves. The important result of the models is that \NHo/$N(\mbox{Si$^{3+}$})\ga10$ for $10^{15}\la N(\mbox{Si$^{3+}$})\la10^{16}$~\cmmt; for lower values of $N(\mbox{Si$^{3+}$})$ the minimal value of the ratio goes up to $\sim50$. When $N(\mbox{Si$^{3+}$})\ga10^{16}$~\cmmt\ the ratio drops a bit below 10, but for that range \NHo$\ga10^{17}$~\cmmt\ and yields a saturated \Lya\ absorption profile. Raising $U$ above 10 does not lower the minimal \NHo/\NSiIV\ below 0.4~dex. This absolute minimum of $\log\NHo/\NSiIV=0.4$ is achieved even for an extreme high value of $\log U=5$. \footnote{Note that $\log U=5$ requires $\log\Sigma=26$ to achieve $\log\mbox{\NSiIV}>17$. This $U$ and $\Sigma$ imply $\tau_{\rm e.sc.}\approx10$, and \cl\ results may be incorrect.}

Using the derived values of \NHo/$N(\mbox{Si$^{3+}$})$, one can set a robust lower limit on $\tau$ of \Lya\ based on the measured $\tau$ of \SiIV. The ratio between the two optical depths is set by the ratio of $f_{12}\lambda$ of the two lines and by \NHo/$N(\mbox{Si$^{3+}$})$. This yields the robust lower limit of
\begin{equation}
\tau(\Lya) \geq 0.71\left[ \frac{N(\mbox{\footnotesize H$^{0}$})}{N(\mbox{\footnotesize Si$^{3+}$})}\right]_{\rm min}\times \tau(\SiIV\mbox{~$\lambda$1394}),\label{eq:tau}
\end{equation}
where the numerical value is calculated using $f_{12}\lambda$ values from \citet{mor91}. Note that [\NHo/$N(\mbox{Si$^{3+}$})]_{\rm min}$ is a function of $N(\mbox{Si$^{3+}$})$. The \Lya\ optical depth is $\sim7\times\tau(\SiIV)$ for the typical value of [\NHo/$N(\mbox{Si$^{3+}$})]_{\rm min}\sim10$. This means that \Lya\ absorption is deeper than \SiIV, unless $\tau\gg1$ and the saturated absorption profile is set by a velocity-dependent covering factor for both lines, and is therefore identical. Note that \citet{kwan90} reached a similar conclusion that $\tau(\Lya) > \tau(\SiIV)$ even while using a steeper slope in the EUV than used here; and while filling the \Lya\ trough by scattered and emitted radiation from other regions.

\subsection{The procedure for setting a robust lower limit on \NHo\ and the absolute metallicity}\label{sec:prescr}
The above described lower limit on $\tau(\Lya)$ for a given $\tau(\SiIV)$ allows to place a robust lower limit on the absorber metallicity according to the following procedure.
\begin{enumerate}
\item Derive the intrinsic $\tau(\SiIV)$ and $N(\mbox{Si$^{3+}$})$ from the observed \SiIV$\lambda\lambda$1394, 1403 absorption profile. The derivation of the intrinsic $\tau(\SiIV)$ depends on the assumed physical properties of the absorber as further described in Sec.~4.3.
\item Find the minimal allowed \NHo/$N(\mbox{Si$^{3+}$})$ by photoionization models and derive the minimal \NHo.
\item Calculate the minimal \Lya\ absorption profile by the absorption system from step (i) using Eq.~\ref{eq:tau}.
\item Reconstruct the unabsorbed spectrum using the minimal \Lya\ absorption i.e., divide the spectrum by the profile from the previous step.
\item If the reconstructed spectrum is consistent with a ``normal'' AGN spectrum, then the gas metallicity is consistent with solar. If an artificial emission feature is produced, continue to the next step.
\item Scale down the minimal allowed \NHo\ from step (ii) by a parameter $s$, and repeat steps (iii)--(iv). Iterate until an ``acceptable'' AGN spectrum is achieved. The lower limit on the absorber metallicity is then $s\times Z_{\sun}$.
\end{enumerate}
This procedure can be trivially extended to cases where several prominent absorption systems are needed to reconstruct the \SiIV\ absorption system. Note that the last step assumes that the minimal allowed \NHo/$N(\mbox{Si$^{3+}$})$ scales linearly with the metal abundance. The assumption certainly breaks for metallicity larger than a few times solar due to the non linear response of the slab ionization structure with increasing metallicity. This does not hinder the identification of objects with super solar metallicity ($Z\ga10Z_{\sun}$). But, placing a correct lower limit on $Z$ for these objects requires a grid of photoionization models with $Z\ga10Z_{\sun}$.

\section{SDSS BALQ sample analysis}\label{sec:data}
We investigate below whether there are any observed BALQs in the SDSS database that present evidence for a high metal/H outflow. Derivation of the BALQ data set, construction of a non-BALQ control sample, spectral analysis and the metal/H estimate are described below.

\subsection{The BALQ data set}\label{sec:balq_data_set}
The data set is drawn from the SDSS DR5 BALQ catalogue of \citet{sca09}. \citet{sca09} constructed the most complete catalogue of BALQs in DR5 (in the $1.7<z<4.2$ rage). They cross-checked the object classification by their method with that of \citet{gibsonetal09}, and visually inspected and classified the objects for which the two classifications disagree. This procedure lowered the probability for a false negative classification from $\sim20$ per cent for the \citet{gibsonetal09} catalogue to $\sim4$ per cent \citep{sca09}. The \citet{sca09} catalogue comprises 3552 objects. The object spectra were retrieved from the SDSS DR7. A successive series of selection criteria is imposed on the spectra. Each criterion is described below in the same order it is imposed. Note that the \citet{sca09} catalogue includes objects with BAL that are not shifted from the emission-line center. Since we want to minimize the uncertainty on the measured absorption profile due to the underlying unknown emission profile, we require a blueshift of at least $\sim1000$~\kms\ of the absorption from the emission-line center. The second criterion described below happens to exclude objects that do not meet this requirement.
\begin{enumerate}
\item We include only objects with $z>2.7$ to ensure that the spectrum contains a sufficiently large ($\sim200$~\AA) wavelength region blue-ward of the \Lya\ emission line. The region is chosen such that it is not affected by the intrinsic \Lya\ absorption. This region is used in the following steps to constrain the intrinsic unabsorbed spectrum. This criterion leaves 973 objects in the data set.\footnote{The redshift of 13 DR5 objects has been changed to values lower than $2.7$ in DR7. Note that while the DR5 redshifts are used to choose objects from the \citet{sca09} catalogue, the revised DR7 redshifts are adopted to calculate the rest-frame wavelengths throughout the analysis.}
\item Since the spectra have relatively low S/N ($\sim5$ per resolution element at continuum regions unaffected by absorption)\footnote{The S/N is estimated by dividing the mean $f_\lambda$ by the standard deviation of $f_\lambda$ in the $\lambda_{\rm rest}=1600-1640$~\AA\ window.}, the spectra are smoothed by a 22 pixel-wide moving-average filter to achieve S/N of at least $\sim15$ per bin [the width is equivalent to $\sim9$ resolution elements \citep{york00}]. We then inspect the spectrum by eye for objects having a \SiIV\ absorption trough with width larger than 15~\AA\ ($\sim3000$~\kms) but smaller than $\sim40$~\AA, to prevent a significant blending between \Lya\ and \NV\ absorption troughs, and normalized depth smaller than 0.5, to select objects with a prominent \SiIV\ absorption. This yields a data set of 139 objects. 
\item The \Lya\ and \SiIV\ wavelength regions are examined by eye to evaluate the S/N within the absorption trough of each spectrum. Objects with poor S/N ($<1$ per bin) in one or both regions are excluded from the analysis, leaving 78 objects. 
\item In this step, we assume that the whole absorption profile is set by a velocity dependent covering factor (CF; i.e., a saturated profile). Note that the assumption yields the minimal possible absorption of \Lya\ (see Sec.~\ref{sec:method}). The \SiIV\ absorption profile is measured. This profile is used to reconstruct the unabsorbed spectrum in the \Lya\ wavelength region. Then, the following procedure is carried out to determine whether the reconstructed spectrum in the \Lya\ region {\it can be} consistent with an unabsorbed \Lya\ emission. A global power-law continuum is fitted for each spectrum using the measured $f_\lambda$ at $\sim1300$ and $1700$~\AA, where $f_\lambda$ is the observed flux density.\footnote{If no featureless window is found around $\sim1300$~\AA, then $f_\lambda$ at $\sim1700$ and $2100$~\AA\ is used. If the 2100~\AA\ is unavailable (due to redshift), then two wavelengths in the region of 1700~\AA\ are used.} Then, a synthetic \Lya\ emission line is placed on top of the continuum fit as an estimate of the intrinsic unabsorbed \Lya\ emission. Since the goal of this analysis is to find objects that require metallicity larger than solar with a high statistical significance, we use a synthetic line that has a large EW of 120~\AA\ and its FWHM is varied between 2000 and 10,000~\kms. The chosen EW corresponds to the upper $\sim15$ per cent of the \Lya+\NV\ blend EW distribution values, derived by \citet{diast09} for $z>3$ non-BALQs. Objects for which the reconstructed spectrum is well above the synthetic emission lines (as determined by eye) are selected. This step leaves 13 objects in the data set. 
\item Spectra of 11 of the objects are similar, having a prominent presence of low-ionization absorption lines (\CII\ and \AlIII) in particular. We chose to analyse only those objects, since they might represent a particular subgroup of BALQ. The other two objects have a \SiIV\ absorption trough shallower than \CIV, and a weak \AlIII\ absorption.
\item To make the sample more homogeneous in terms of $z$ and $M_{\it i}$ values, we exclude 3 more objects. The homogeneity is important in order to allow a comparison to non-BALQs (see below). We use $M_{\it i}$ from \citet{sch10} as a proxy for the luminosity, where {\it i} is the SDSS {\it i} filter.\footnote{The values used are from the VizieR catalog. They do not appear in the original paper of \citet{sch10}.} Two of the objects have a relatively high $z$ ($\sim3.8$) and one object has a high absolute magnitude ($M_{\it i}\sim-28.7$) and are therefore excluded for the sake of homogeneity. The final data set is comprised of 8 objects with $2.70<z<3.07$ and $-27.72<M_{\it i}<-26.18$. 
\end{enumerate}
Table~\ref{tab:all_objects} lists the name, $z$ and $M_{\it i}$ of the final data set objects. Note that no conclusion with a statistical meaning is drawn from the few objects in the final data set. The goal of the analysis is to check whether there are objects consistent with a high metal/H outflow.

\begin{figure*}
\includegraphics[width=174mm]{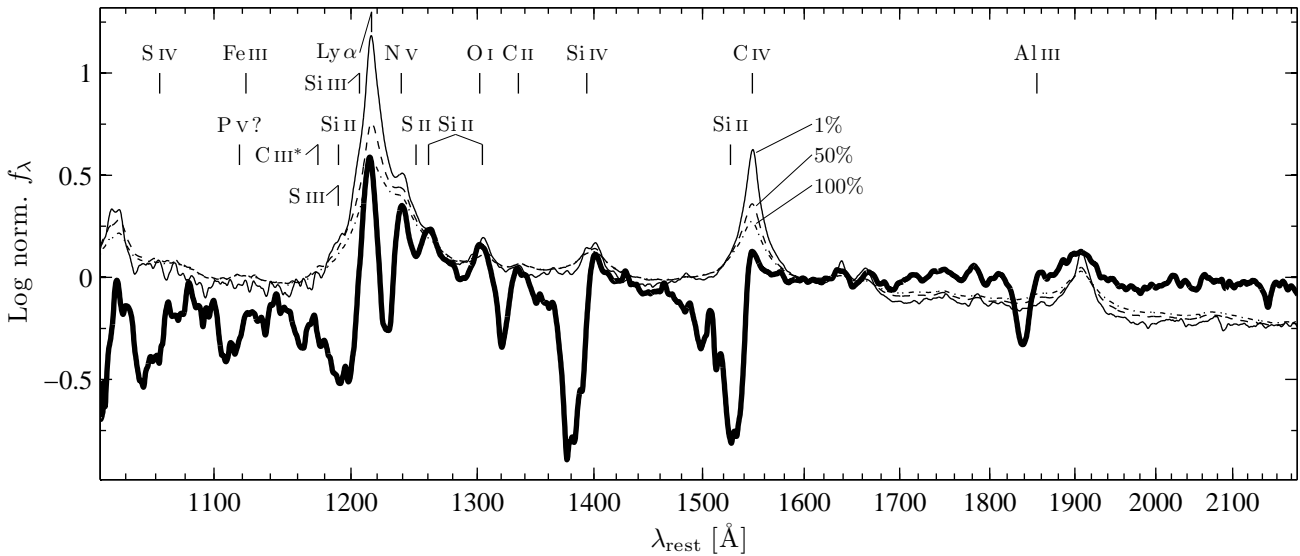}
\caption{Composite spectra of the 8 BALQ sample (thick line) and control sub-sample composites: the top 1 per cent (thin line), the top 50 per cent (dashed line), and the whole control sample (i.e., 100 per cent; dash-dotted line). All spectra are normalized using a mean $f_\lambda$ in the range of $\sim1600-1640$~\AA. The laboratory wavelengths of the bluest line of each multiplet are indicated (the transition to the ground-state is indicated for \CII). Note the presence of prominent absorption features of low-ionization lines, in particular \CII\ and \AlIII; and the similarity between absorption profiles of \SiIV\ and \CIV\ lines.}\label{fig:composite_plot}
\end{figure*}

Figure~\ref{fig:composite_plot} presents a composite spectrum of the 8 objects of the data set. Each spectrum is smoothed as described above [step (ii)] and normalized using a mean $f_\lambda$ in the wavelength window of $\sim1600-1640$~\AA. The composite spectrum is a geometric mean of the smoothed and normalized spectra. The similarity of \SiIV\ and \CIV\ absorption profiles  strongly suggests $\tau\ga1$ and similar covering factor for both lines. Note the almost complete absence of \CIV\ in emission. This is unlikely to be a result of absorption because the red-wing of the emission line appears smooth and without any features that can be attributed to absorption. The \Lya\ emission line appears also weak but this might be attributed to an absorption of the red wing by \NV\ and of the blue wing by \Lya. Note also the prominent absorption features of the relatively low-ionization lines, in particular \CII\ and \AlIII.

\begin{table}
\begin{minipage}{84mm}
\caption{Parameters of the 8 objects that comprise the final BALQ data set.}\label{tab:all_objects}
\begin{tabular}{@{}{l}*{2}{c}@{}}
\hline
Name (SDSS J) & $z$ & $M_{\it i}$\fnrepeat{fn2:m_i}\\
\hline
120954.14+142319.7  & 2.7005 & $-26.364$ \\
145138.29+415401.0 & 2.7425 & $-26.908$ \\
134934.14+245540.1  & 2.7706 & $-27.037$ \\
143506.16+240144.8 & 2.8096 & $-27.717$\\
125224.35+144508.7 & 2.9048 & $-26.185$\\
111748.56+392746.2 & 2.9125 & $-27.578$\\
102321.90+493936.4 & 3.0342 & $-26.922$ \\
134818.03+423205.1 & 3.0661  & $-26.938$ \\
\hline 
\end{tabular}
\footnotetext[1]{The absolute magnitude in the SDSS {\it i} filter. Values are from \citet{sch10}.\label{fn2:m_i}}
\end{minipage}
\end{table}

\subsection{The non-BALQ control sample}
The procedure presented above requires a control non-BALQ spectrum to estimate the unabsorbed \Lya\ profile. We draw the control sample of non-BALQs from the SDSS DR7 quasar catalogue of \citet{sch10}. We select objects having values of $z$ and \Mi\ in the range of the BALQ data set (see above). We exclude objects that are marked as BALQ in at least one of the catalogues of \citet{sca09}, \citet{allenetal11} and \citet{shenetal11}. This comprises a sample of 1845 objects. Individual spectra are smoothed by a 22 pixel-wide moving-average filter and normalized using a mean $f_\lambda$ in the range of $\sim1600-1640$~\AA\ (the same procedure as for the BALQ data set). The non-BALQ objects are sorted by the \Lya\ emission strength. The mean normalized $f_\lambda$ in a window of $1213.6-1227.8$~\AA, $f_\lambda(\Lya)$, is utilized as a proxy for the \Lya\ emission strength (the wavelength window is equivalent to $-500<v<3000$~\kms\ relative to \Lya~$\lambda1215.67$). We choose the lower wavelength boundary to reduce the effect of intervening \Lya\ absorption. The upper boundary is chosen to reduce the contamination from the \NV\ emission line. Objects having mean $f_\lambda(\Lya)<1$ (i.e., \Lya\ emission peak is lower than the continuum at $\sim1620$~\AA), or S/N in the {\it i} filter smaller than 3 are excluded from the sample. This criterion excludes 73 objects, yielding a control sample that contains 1772 objects.

The non-BALQ objects are divided into sub-samples based on the \Lya\ strength. The first sub-sample contains the top 1 per cent of the objects (18 objects) in terms of the \Lya\ strength, the second -- 5 per cent, and the successive sub-samples contain 10, 20 percent, etc.\ of the objects. A geometric mean spectrum is calculated for each sub-sample and for the complete control sample. Figure~\ref{fig:composite_plot} presents composite spectra of the 1 and 50 per cent sub-samples and of the complete sample. The composite spectra are overall similar, except in the strength of \Lya\ and \CIV\ emission lines. In particular, there is a similarity in the continuum level blue-ward of \Lya\ which appears not to be significantly affected by intervening \Lya\ absorption. Note that the BALQ composite spectrum is redder than the control composite spectra (e.g., \citealt{reichardetal03}).

\subsection{Estimate of the metallicity}
The structure of this section is as follows. We begin by an outline of how different physical absorption settings can affect the measured column density (Sec.~\ref{sec:why_cases}). This is followed by a description of the column density measurement procedure for four assumed physical settings (Secs.~\ref{sec:case_i} -- \ref{sec:case_iv}). Then, the procedures for estimating the normalized absorption profiles and the outflow metallicity are described (Secs.~\ref{sec:norm_abs} and \ref{sec:est_proced}, respectively). Finally, the direct lower limits on absolute metallicity of the BALQ-sample objects are presented (Sec.~\ref{sec:est_results}).

\subsubsection{What sets the measured column density?}\label{sec:why_cases}
The measured column density critically depends on the assumed physical setting of the absorber. If the profile is set by a foreground uniform absorber with a constant covering factor in velocity space, then the profile is governed by $\tau$ only and the column density can be readily measured. If the profile is set by a velocity dependent CF, then $\tau\gg1$ is possible and the column density cannot be constrained. There can also be a combination of these two extreme settings. Thus, we have assumed several different physical settings of the absorbing gas and measured the resulting intrinsic $\tau(\SiIV)$ for each of the 8 objects in the BALQ date set. The doublet nature of the \SiIV\ line can be utilized to rule out the $\tau\gg1$ setting for absorption profiles that have a deep dip in the blue part of the profile without a similar dip in the red part. The intrinsic $\tau(\SiIV)$ is an output of the first step of the metallicity estimate procedure described in Sec.~\ref{sec:prescr}. The different physical setting cases are described below.

\subsubsection{Case (i) -- Non saturated absorber with a constant CF and a $N(v)$ with a Gaussian profile}\label{sec:case_i}
The absorption profile is mostly set by \Nion\ as a function of $v$ and we assume a Gaussian velocity distribution with a standard deviation $b$. Note that our goal is to fit the main bulk of the absorption rather than the exact profile with its small features. We follow the method to estimate the intrinsic $\tau(\SiIV)$ that is described in \citet{bl08}, and is briefly reviewed here. Each intrinsic synthetic absorption system has four free parameters: the velocity shift $v_{\rm shift}$, Doppler broadening parameter $b$, global CF, and the ionic column density \Nion. The first three parameters are mostly set by the overall shape of the absorption profile. The ionic column density is determined by the intensity of the absorption, but it has some degeneracy with the global CF. The intrinsic synthetic profile is transformed to an observed synthetic profile by convolving the former with the same smoothing function as the spectra. We do not convolve the intrinsic synthetic profile with the line-spread function (LSF) of the instrument, since the LSF width ($\sim170$~\kms) is significantly smaller than the width of the absorption ($\ga3000$~\kms). The observed spectrum is corrected by the synthetic observed absorption profile that is produced by assuming a set of free parameters. The set is varied until the correction produces a reasonable reconstruction of a continuum and line emission. The success of the reconstruction is determined by eye. The \Lya\ profile is then corrected using the resulting $b$, CF and $N(\mbox{Si$^{3+}$})$, and the equality in Eq.~\ref{eq:tau}, where the [\NHo/$N(\mbox{Si$^{3+}$})]_{\rm min}$ is derived at the estimated $N(\mbox{Si$^{3+}$})$.

We prefer to use case (i) as the prime one. This is because the fitting method used for this case, although not robust, is more certain than the other methods (described below). These methods require an {\it a priori} estimate of the unabsorbed spectrum in regions affected by emission lines, whose shape is not well constrained. We use cases (ii)--(iv) only for objects for which case (i) yields a lower limit on the metallicity that is higher than solar, since we want to check whether these cases also require $Z>Z_{\sun}$.

\subsubsection{Case (ii) -- Non saturated absorber with a constant CF and a $N(v)$ with an arbitrary profile}\label{sec:case_ii}
The absorber is modelled by $\tau(\SiIV;v)$ and a global CF. The difference between this method and the previous one is that $\tau(\SiIV;v)$ is calculated directly at each velocity bin, which we chose to be the instrumental width ($\sim70$~\kms). We do not assume that $N(\mbox{Si$^{3+}$}; v)$ has a Gaussian profile, but rather calculate it directly at each velocity bin assuming thermal broadening that is smaller than the bin width (i.e., $b\sim10$~\kms). The column density at each velocity bin is calculated through the relation $\tau(v)=\frac{\pi e^2}{m_e c}f_{12}\lambda_{\rm ion}\,dN(v)/dv$ [hereafter, $N(v)\equiv dN(v)/dv\times\Delta v$, where $\Delta v$ is the instrument bin width]. The main difficulty to execute this procedure is the blending of the \SiIV~$\lambda\lambda$1394, 1403 doublet. One needs to de-blend the contribution of the two lines at a given $\lambda$. \citet*{junkkaetal83} suggested to overcome this difficulty by measuring the absorption profile beginning from its bluest part, where only the \SiIV~$\lambda$1394 contributes to the absorption. Then, using this measurement and the ratio of $f_{12}\lambda$ of the two \SiIV\ lines, the expected \SiIV~$\lambda$1403 absorption is calculated at a velocity that is shifted by the velocity difference between the two lines. If the observed absorption at this velocity is larger than the expected \SiIV~$\lambda$1403 absorption, an additional \SiIV~$\lambda$1394 absorption is added at that velocity. The procedure is carried out throughout the whole absorption profile. Note that \citet{junkkaetal83} did not include effects of CF in their method. These effects are accounted for in this study. 

The contribution of the blue and red line of the \SiIV\ doublet is de-blended as follows. The calculation of $N(\mbox{Si$^{3+}$};v)$ begins from the bluest part of the absorption profile, assuming a contribution only from the \SiIV~$\lambda$1394 line. We use the relation between the normalized absorption profile $I$ and $\tau$, $I(v)=1-{\rm CF}+{\rm CF}\exp[-\tau(v)]$ (the determination of $I(v)$ is described in Sec.~\ref{sec:norm_abs}). This relation is utilized until the separation between \SiIV~$\lambda$1394 and \SiIV~$\lambda$1403 in velocity scale is reached ($\sim1940$~\kms). Then, $N(\mbox{Si$^{3+}$};v)$ is calculated by the relation 
\[
I(v)=1-{\rm CF}+{\rm CF}\exp[-\tau(v)]\times\exp[-\tau(v-1940)/2].
\]

The \Lya\ profile is calculated using the calculated $\tau(\SiIV\mbox{~$\lambda1394$};v)$ and Eq.~\ref{eq:tau}, where [\NHo/$N(\mbox{Si$^{3+}$})]_{\rm min}$ is estimated assuming two cases. First, the whole outflow is treated as a single photoionized slab, e.g. the outflow is photoionized by photons entering it at a surface where $v=0$. Second, each velocity bin is assumed to represent a separate photoionized slab e.g., a clumpy outflow with clumps that do not cover the ionization source from each other (i.e., no self coverage). The latter case produces higher metallicity because [\NHo/$N(\mbox{Si$^{3+}$})]_{\rm min}$ is larger for lower $N(\mbox{Si$^{3+}$})$ (see Fig.~\ref{fig:ion_struc2}). Intermediate cases produce intermediate metallicity. Note that the derived metallicity is almost insensitive to the precise choice of the velocity bin width $\Delta v$, because the underlying relation used in the procedure is between $\tau(\Lya)$ and $\tau(\SiIV)$.

\subsubsection{Case (iii) -- A velocity dependent CF along the line of sight}\label{sec:case_iii}
In this case, we assume that the absorber acceleration is along our line of sight to the emission source, and that the absorption profile is mainly set by a velocity dependent CF. We do not try to measure $\tau(\SiIV)$ but rather estimate its lower limit. The measurement of CF$(v)$ and $\tau(\SiIV;v)$ is conducted iteratively. First, we assume $\tau(v)=\infty$ and calculated CF$(v)$ through $I(v)=1-{\rm CF}(v)+{\rm CF}(v)\exp[-\tau(v)]$ beginning from the bluest part of the absorption profile. When the separation between \SiIV~$\lambda$1394 and \SiIV~$\lambda$1403 is reached, we introduce the complementary part into the relation 
\[
\begin{array}{rcl}
I(v)&=&\{1-{\rm CF}(v)+{\rm CF}(v)\exp[-\tau(v)]\}\\
& & \times\{1-{\rm CF}(v^\prime)+{\rm CF}(v^\prime)\exp[-\tau(v^\prime)/2]\},
\end{array}
\]
where $v^\prime=v-1940$~\kms. If no physical solution to ${\rm CF}(v)$ at a given $v$ can be found (i.e., the predicted absorption by \SiIV~$\lambda$1403 assuming $\tau\gg1$ is larger than the observed one), we assume for this iteration step ${\rm CF}(v)=0$ (i.e., \SiIV~$\lambda$1394 does not contribute to the absorption at this $v$) and calculate $\tau(v^\prime)$, which is then used as an input for the next iteration step. The process is iterated until a physical solution to ${\rm CF}(v)$ is found for all $v$, or when the maximum number of iterations is reached. The \Lya\ profile is estimated using ${\rm CF}(v)$, $\tau(\SiIV\mbox{~$\lambda1394$};v)$ and Eq.~\ref{eq:tau}. Here we assume the minimal value of [\NHo/$N(\mbox{Si$^{3+}$})]_{\rm min}\approx2.5$ (see Fig.~\ref{fig:ion_struc2}). Note that this case also checks whether the profile can be fitted by ${\rm CF}(v)$ and $\tau\gg1$. If it can, the algorithm converges after the first iteration.

\subsubsection{Case (iv) -- A velocity dependent CF perpendicular to the line of sight}\label{sec:case_iv}
In this case, we assume that the absorber acceleration is perpendicular to our line of sight. The procedure is similar to the previous case, except we use the following equation to deblend the profile:
\[
\begin{array}{rcl}
I(v)&=& 1-{\rm CF}(v)+{\rm CF}(v)\exp[-\tau(v)]\\
& & -{\rm CF}(v^\prime)+{\rm CF}(v^\prime)\exp[-\tau(v^\prime)/2].
\end{array}
\]
This case also checks whether the profile can be fitted by ${\rm CF}(v)$ and $\tau\gg1$. 

\subsubsection{Estimating the normalized absorption profile}\label{sec:norm_abs}
An estimate of a normalized \SiIV\ absorption profile $I$ (i.e., absorbed flux divided by the estimated unabsorbed flux) is calculated in order to apply procedures in cases (ii)-(iv) described above. We approximate the unabsorbed spectrum in the \SiIV\ region by a constant value. The value is set by a mean $f_\lambda$ at the same range that is used to normalize the spectra ($\sim1600-1640$~\AA). This provides a good approximation, as determined visually, for objects requiring the use of cases (ii)-(iv). This treatment under-predicts the absorption in the red wing of the profile, as it does not take into account absorption of the \SiIV\ emission. Thus, the implied \Lya\ absorption is also underpredicted.

\subsubsection{The metallicity estimate procedure}\label{sec:est_proced}
The minimal required $N(\Ho)$, implied by the measured $N(\mbox{Si$^{3+}$})$,
is derived using \cl\ models with solar metallicity (see Sec.~3). A curve of $\Sigma$ versus $U$ is calculated for the measured $N(\mbox{Si$^{3+}$})$ (e.g., Fig.~\ref{fig:ion_struc2}). The \cl\ models are interpolated using a 0.1~dex grid in $\Sigma$ and $U$. The curve is utilized to derive the allowable range of $N(\Ho)$ values. The minimal possible $N(\Ho)$ is then found and used to reconstruct the observed spectrum in the \Lya\ region. The photoionization models of a static slab should describe to a good approximation the photoionization structure of an outflowing gas. The observed outflow velocities of $v\la0.03c$ do not alter considerably the ionizing flux when the latter is transformed to the gas rest-frame.

The non-BALQ control sample is utilized to estimate whether the reconstructed spectrum in the \Lya\ region, using solar metallicity, is consistent with a \Lya\ emission profile; and if it is not -- to give a rough estimate of a lower limit on the metallicity. The excess EW of the reconstructed spectrum is calculated relative to each control sub-sample in the region blue-ward of \Lya. An excess EW $>10$~\AA\ is considered as unacceptable. The value of 10~\AA\ is based on a rough estimate of the associated systematic uncertainties in this procedure. If the reconstructed spectrum is inconsistent with the top 1 per cent sub-sample, the minimal \NHo\ is scaled down by a given factor in the $1.5-100$ range, and the fit procedure is repeated. The maximal factor which still produces an excess ${\rm EW} >10$~\AA\ is adopted as the object $Z/Z_{\sun}$.

\subsubsection{Metallicity estimate results}\label{sec:est_results}
Four objects are consistent with solar metallicity, assuming case (i): SDSS J120954.14+142319.7, 145138.29+415401.0, 143506.16+240144.8 and 111748.56+392746.2. One object, SDSS J125224.35+144508.7, has an observed spectrum in the blue-wing region of \Lya, where the flux density is already larger than the top object of the control sample prior to making any correction. Thus, the object is excluded from the analysis. 

Three objects are inconsistent with solar metallicity assuming case (i). We check whether the other cases allow lower metallicity than case (i). Cases (iii) and (iv) imply the same lowest lower limit on $Z/Z_{\sun}$ for these objects. We present for each object the case that yields the lowest excess EW relative to the top 1 per cent control sub-sample for $Z=Z_{\sun}$. Table~\ref{tab:measured_param_su_solar} lists the parameters used in case (i) procedure, the resulting excess EW relative to the top 1 per cent control sub-sample, and an estimate of the lower limit of $Z/Z_{\sun}$. The table also lists the $Z/Z_{\sun}$ for cases (iii) and (iv), the case that yields the lowest excess EW and its value.

\begin{figure}
\includegraphics[angle=-90,width=88mm]{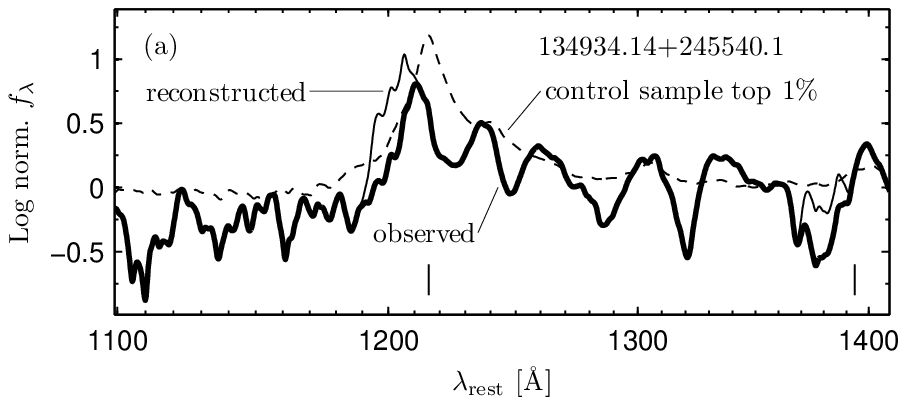}
\includegraphics[angle=-90,width=87mm]{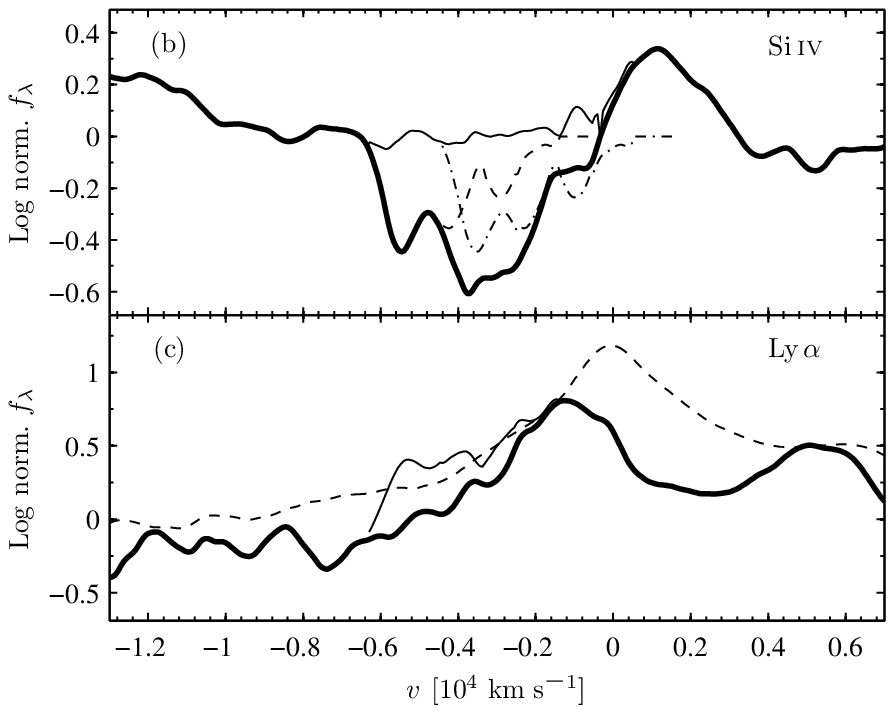}
\caption{Spectrum reconstruction and absorption profile measurement for SDSS J134934.14+245540.1 assuming cases (i) and (iii). Panel (a) presents the reconstructed spectrum assuming case (i). The laboratory wavelength of the blue line of \Lya\ and \SiIV\ doublets is indicated by a vertical tick mark. If the reconstructed spectrum is assumed to represent the intrinsic unabsorbed spectrum, then a very broad and blue-shifted \Lya\ emission line is introduced. Panel (b) presents the measurement of the intrinsic absorption profile assuming a velocity-dependent CF. Procedure of case (iii) is carried out (see text) on the observed normalized profile (thick solid line), yielding an estimate of the intrinsic \SiIV~$\lambda$1394 absorption profile (dashed line). The absorption profile of \SiIV~$\lambda$1403 (dot-dashed line) is produced by shifting the \SiIV~$\lambda$1394 profile by $\sim1940$~\kms. The reconstructed spectrum using the total \SiIV\ absorption is presented (thin solid line). The unabsorbed emission is approximated by a constant ($\log f_\lambda=0$). Note that the \SiIV\ absorption can be fitted assuming $\tau(v)\gg1$ for all $v$. Panel (c) presents the reconstruction of \Lya. The observed spectrum (thick solid line) is corrected by the measured profile assuming solar metallicity, producing an estimate of the unabsorbed spectrum (thin solid line) which is compared to the top 1 per cent control sub-sample (dashed line). The two latter spectra are consistent i.e., the excess EW $<10$~\AA.}\label{fig:fitted_134934}
\end{figure}

Figure~\ref{fig:fitted_134934} presents the results of cases (i) and (iii) measurement procedures for SDSS J134934.14+245540.1 . Case (i) is presented in panel (a). The \Lya\ reconstructed spectrum is compared to the top 1 per cent control sub-sample. The reconstruction produces a blue-shifted \Lya\ emission profile which is inconsistent with the observations of non-BALQs. The measurement of the intrinsic absorption profile using the case (iii) procedure is presented in panel (b) of Fig.~\ref{fig:fitted_134934}, and the reconstructed spectrum in the \Lya\ region is presented in panel (c). SDSS J134934.14+245540.1 is consistent with $\tau(v)\gg1$ for all $v$ i.e., its absorption profile can be completely fitted by CF($v$). Case (iii) implies metallicity consistent with solar for this object.

\begin{figure}
\includegraphics[angle=-90,width=88mm]{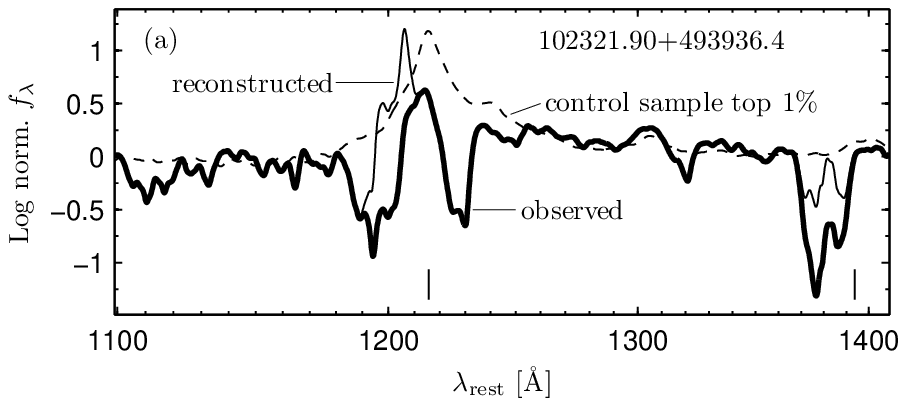}
\includegraphics[angle=-90,width=87mm]{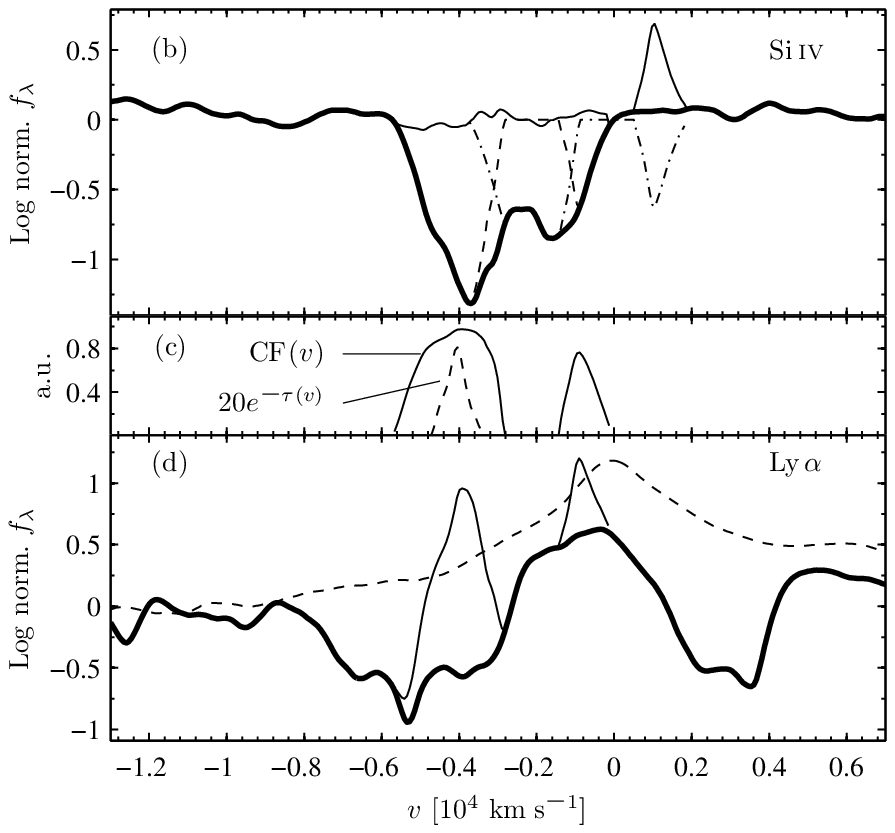}
\caption{Same as Fig.~\ref{fig:fitted_134934} for SDSS J102321.90+493936.4 assuming cases (i) and (iii). The intrinsic ${\rm CF}(v)$ (solid line) and $\exp[-\tau(v)]$ (dashed line) of \SiIV~$\lambda$1394 are presented in panel (c). Note that the \SiIV~$\lambda$1403 component predicts an absorption at $v\sim1000$~\kms\ (an emission feature in the reconstructed spectrum). This is expected because the normalized absorption spectrum is calculated disregarding the expected intrinsic \SiIV\ emission. The \Lya\ reconstructed spectrum [panel (d)] is inconsistent with the top 1 per cent control sub-sample (an excess EW of 27~\AA). Thus, indicating that the metallicity should be higher than solar.}\label{fig:fitted_102321}
\end{figure}

\begin{figure}
\includegraphics[angle=-90,width=88mm]{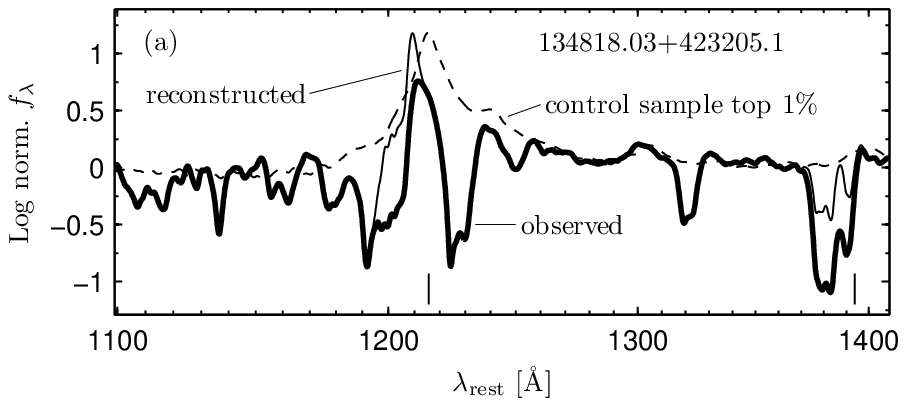}
\includegraphics[angle=-90,width=87mm]{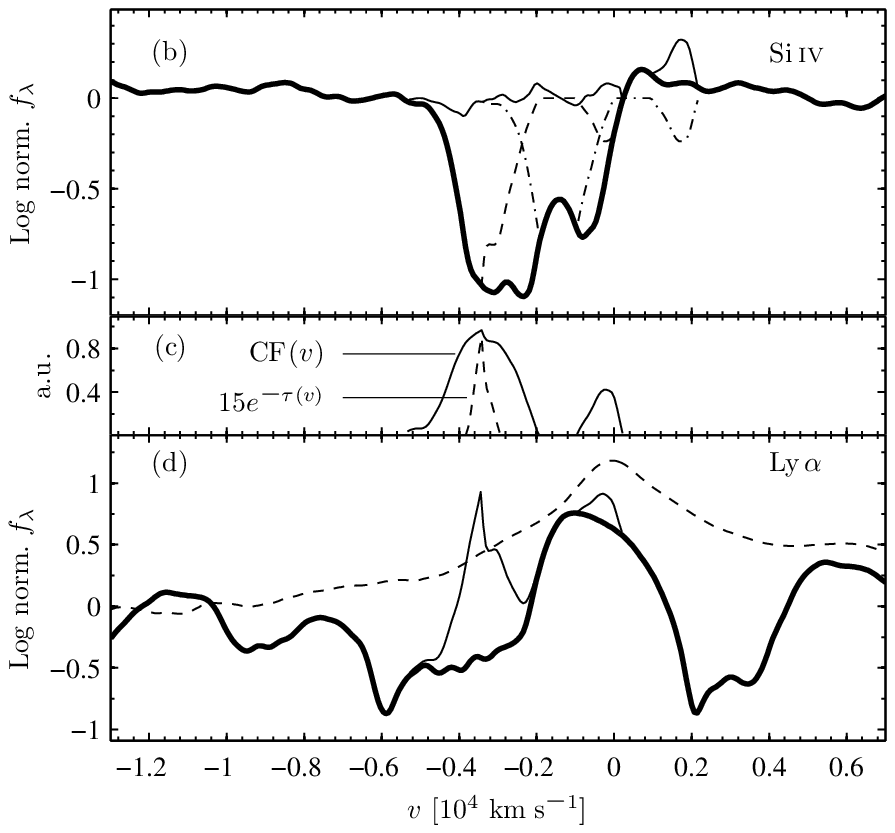}
\caption{Same as Fig.~\ref{fig:fitted_102321} for SDSS J134818.03+423205.1 assuming cases (i) and (iv). The resulting reconstructed \Lya\ spectrum assuming case (iv) is consistent with the top 1 per cent control sub-sample [panel (d)]; i.e., the spectrum is consistent with solar metallicity.}\label{fig:fitted_134818}
\end{figure}

Assuming $\tau(v)\gg1$ for the remaining two objects produces \SiIV~$\lambda$1403 absorption that is significantly stronger than observed. This requires to adopt a finite $\tau(v)$ for cases (iii) and (iv). Figure~\ref{fig:fitted_102321} presents the results of the case (i) and case (iii) measurement procedures for SDSS J102321.90+493936.4. The minimal CF($v$) and the resulting $\tau(v)$ are presented in panel (c). The spectrum of SDSS J102321.90+493936.4 remains inconsistent with solar metallicity for all cases. The lower limit for this object is $Z \ga2Z_{\sun}$ from case (iii) measurement. Figure~\ref{fig:fitted_134818} presents the results of cases (i) and (iv) for SDSS J134818.03+423205.1. Case (iv) applied to this object yields a metallicity that is consistent with solar. Thus, only the absorption is SDSS J102321.90+493936.4 yields a robust result for higher than solar metallicity.

\begin{table*}
\begin{minipage}{150mm}
\caption{Parameters of the absorption system for the BALQ data set objects which are inconsistent with solar metallicity assuming case (i), and the lowest metallicity achieved assuming cases (ii)--(iv).}\label{tab:measured_param_su_solar}
\begin{tabular}{@{}l*{11}{c}@{}}
\hline
 Name (SDSS J) & $v_{\rm shift}$ & $b$\fnrepeat{fn5:b} & CF & $N(\mbox{Si$^{3+}$})$\fnrepeat{fn5:ion1} & \NHo\fnrepeat{fn5:ion2} & EW\fnrepeat{fn5:ew} & $Z$\fnrepeat{fn5:z} & Case\fnrepeat{fn5:case} &EW\fnrepeat{fn5:ew2} & $Z$\fnrepeat{fn5:z2}\\
 &  (\kms) & (\kms) & & & & (\AA) & & & (\AA) & \\
\hline
134934.14+245540.1 & $-3800$ & 1000 & 0.70 & 15.34 & 16.76 & 49 & 18 & (iii) & \ 7 & 1\\
102321.90+493936.4 & $-3700$ & 1000 & 0.90 & 15.48 & 16.83 & 39 & \ 5 & (iii) & 27 & 2\\
134818.03+423205.1 & $-3000$ & \ 900 & 0.80 & 15.65 & 16.87 & 26 & \ 4 & (iv) &\ 5 & 1\\
\hline
\end{tabular}
\footnotetext[1]{Doppler broadening parameter.\label{fn5:b}}
\footnotetext[2]{Log of the ionic column density in units of cm$^{-2}$.\label{fn5:ion1}}
\footnotetext[3]{Log of the \emph{minimal} H$^0$ column density in units of cm$^{-2}$.\label{fn5:ion2}}
\footnotetext[4]{The excess EW relative to the top 1 per cent control sub-sample for case (i) and solar metallicity.\label{fn5:ew}}
\footnotetext[5]{Lower limit on the metallicity in units of $Z_{\sun}$ for case (i).\label{fn5:z}}
\footnotetext[6]{Both cases (iii) and (iv) imply the same lower limit on $Z$. The case that yields the lowest excess EW is indicated.\label{fn5:case}}
\footnotetext[7]{The excess EW relative to the top 1 per cent control sub-sample for the indicated case and solar metallicity.\label{fn5:ew2}}
\footnotetext[8]{Lower limit on the metallicity in units of $Z_{\sun}$ for the indicated case.\label{fn5:z2}}
\end{minipage}
\end{table*}

\section{Discussion}
A mechanism which can produce outflows with a high metal/H abundance ratio is described in Sec.~\ref{sec:metal}. Radiation pressure is mainly exerted on metals, which are coupled to protons and electrons by the Coulomb force. For a sufficiently large flux of photons, the acceleration of metals by radiation pressure is larger than the deceleration due to Coulomb friction, and the metals can decouple (i.e., run away) from the H gas, and produce a pure metal outflow. Several observed phenomena that may be explained by the proposed mechanism are describe below (Sec.~\ref{sec:dis_obs_phe}). We also discuss possible direct observational evidence for this mechanism (Sec.~\ref{sec:dis_dirct_evi}), and regions of the active nucleus that are suitable for the metal runaway mechanism (Sec.~\ref{sec:dis_where}). Finally, several caveats to the proposed metal separation mechanism are described (Sec.~\ref{sec:dis_cav}).

\subsection{Extreme high velocity outflows}\label{sec:dis_obs_phe}
The metal separation mechanism may be relevant to some observations which at face-value are inconsistent with the radiation-driven outflow scenario. It may explain the extremely high outflow velocities, $v\ga 50,000$~km~s$^{-1}$, found by \citet{jannuzietal96}, \citet{rhidalgoetal07} and \citet*{rhidalgoetal11}, where one of the objects, PG~0935+417, appears to have a relatively modest $L/L_{\rm Edd}\sim0.2$ \citep{rhidalgoetal11}.\footnote{The ionic column densities reported by \citet{rhidalgoetal11} for PG~0935+417 cannot be utilized to constrain the object metallicity by using the method described in Sec.~\ref{sec:method}. The authors provide only upper limits on \Nion\ for the low ionization species [including \NHo\ that is estimated using the \Lya\ and \Lyb\ absorption lines] and for $N(\mbox{N$^{4+}$})$. The \CIV\ absorption cannot be used to constrain the metallicity, since the object shows variable absorption \citep{rhidalgoetal11} and \CIV\ was not observed simultaneously with \Lya\ and \Lyb. The remaining absorption line is \OVI, which is a high ionization specie line (requires ionizing photons with $h\nu>8.4$~Ryd) and implies a very modest absolute lower limit of $\NHo=0.02N(\mbox{O$^{5+}$})$ for $\log N(\mbox{O$^{5+}$})\approx15-17$. The reported lower limit on $N(\mbox{O$^{5+}$})\ga8.2\times10^{15}$~\cmmt\ results in $\NHo\ga0.2\times10^{15}$~\cmmt\ which is within the reported upper limit on \NHo. Note that the measurement of all absorption lines except \CIV\ is hindered by the intervening \Lya\ forest for this object.} The terminal velocity $v_{\rm term}$ of a radiation-driven outflow is $v_{\rm term}\approx\sqrt{\mathcal{M}\times L/L_{\rm Edd}}v_{\rm esc}$, where $\mathcal{M}$ is the force multiplier and $v_{\rm esc}=\sqrt{2}v_{\rm Kep}$ is the escape velocity at the origin of the outflow, where $v_{\rm Kep}$ is the Keplerian velocity (e.g., \citealt{chenet01,laorbrandt02}). The radiation pressure is exerted on the metals, which constitute only $\sim 1$ per cent of the total mass for $Z=Z_{\sun}$. A higher metal/H outflow implies the metals drag with them less H, and thus $\mathcal{M}\propto Z$, and therefore  $v_{\rm term}\propto \sqrt{Z}$. Since $Z_{\sun}\sim 0.01$, a pure metal outflow i.e., with $Z\sim 1$, will reach $v_{\rm term}$ larger by a factor of 10. Only $\sim 1$ per cent of the total mass, which resides in metals, is being accelerated. Since the radiation pressure is unchanged, the acceleration is 100 times larger, and for the same acceleration region size, the final speed is 10 times larger.

The runaway mechanism may also explain the report by \citet{cottisetal10} of the apparent lack of \Lya-\NV\ line-locking signature in the \CIV\ absorption trough in objects that meet all of the five physically motivated criteria to possess this signature, as
suggested by Arav (1996). The high metal/H outflow most likely can be approximated as a multi-component fluid with a weak coupling between different components. The weak coupling is due to the relatively low number density of the fluid constituents. An ion runaway will cause a drop to $\sim10^{-3}$ in the particle number density, compared to a solar abundance gas. The coupling between an ion and an ion-fluid due to Coulomb scattering is $\sim Z_{\rm ion,fl}^2$ ($\sim4^2\sim20$) times larger than that between an ion and a H-fluid, where $Z_{\rm ion,fl}$ is the charge of the ion-fluid (note that $Z^2_{\rm ion,fl}=1$ in Eq.~\ref{eq:fric}, since the equation is written for an ambient fluid composed from protons and electrons only). This gives a $\sim1/(10^{-3}\times20)=50$ drop in the coupling force of an ion to an ion-fluid. Thus, the coupling between C and N is negligible in the outflow, and no \Lya-\NV\ line-locking is expected for the C absorption lines.

A direct observational manifestation of the weak coupling between different ions of a high metal/H outflow can be a difference in the absorption velocity distribution of different ions, which cannot be simply explained by a change of ionization with the outflow velocity. Since, $v_{\rm term}\propto\sqrt{\mathcal{M}}$, $\mathcal{M}\propto f_\nu(\lambda_{\rm ion}) f_{12}/m_{\rm ion}$ (Eq.~\ref{eq:a_rad}). Assuming the same illuminating point source with $f_\nu\propto\nu^{-\alpha}$ for all ions, yields $v_{\rm term}({\rm ion})\propto \sqrt{\nu_{\rm ion}^{-\alpha}f_{12}/m_{\rm ion}}$, and one needs to sum up all resonance transitions contributing to the acceleration of a given ion. It may be difficult to observationally identify relatively narrow extreme high velocity outflows, if different ions reach different $v_{\rm term}$. The current technique to identify such systems is by looking for similar velocity shifts of the narrow absorption profiles, generally of \CIV, \NV\ and \OVI, as for example measured in PG~0935+417 \citep{rhidalgoetal11}.

\subsection{The outflow metallicity}\label{sec:dis_dirct_evi}
We do not find robust evidence for a metal runaway (say by finding $Z\ga10Z_{\sun}$) in the objects analysed here. The measured metallicities are similar to the absolute metallicities found in earlier studies of BALQs \citep{koristaetal96} and NALQs \citep{hamannetal97, aravetal07, hamannetal11}. The moderate metal over abundance ($Z\ga2Z_{\sun}$) may reflect metal enrichment produced by enhanced star formation rate close to the active nucleus.

Utilizing relative metal abundances to test the runaway scenario is not straightforward. The relative metal abundances in the metal runaway scenario depend on both the initial metal enrichment, and the decoupling and acceleration of the different ions. Observed outflows with over-abundance of some metals relative to other (e.g., \citealt{wangetal09}) may be caused by an SED which allows only the over-abundant metals to run away efficiently. In the local Universe there is evidence for both sub- and super-solar abundances in galaxies (\citealt*{zaritskyetal94}; \citealt{gallazzietal05}), observations at $z>1.5$ point towards sub-solar abundances in central regions of galaxies (\citealt{crescietal10, jonesetal10, yuanetal11}), and the high metal/H outflow may be characterized by a low metallicity metal abundance ratio.

\subsubsection{Prospects of future observations of nearby BALQs}\label{sec:future}
A high quality spectrum of the UV rest frame of nearby BALQs ($z\la1$) will allow to place a better constraint on the outflow absolute metallicity, as lower $z$ AGN are significantly less absorbed by intervening \Lya\ absorption systems. The detected \Lya\ absorption will reflect the intrinsic \Lya\ absorption only, and will allow an accurate comparison of the relative \Lya\ versus \SiIV\ absorption profiles. Also, extreme high velocity metal systems will be detectable without confusion by the \Lya\ forest, in particular since such systems tend to be narrow.

\subsection{Where can the high metal/H outflow originate from?}\label{sec:dis_where}
There are two basic conditions for producing a high metal/H outflow. First, a metal runaway can be produced only if there is sufficient photon flux relative to the H density in a wavelength region that is absorbed by metal absorption lines e.g., $\log f_\nu/n_{\rm H}\ga-10.2$ for $T=10^4$~K gas (see Eq.~\ref{eq:runaway}), which yields $\log U\ga5$ adopting the \cl\ standard AGN SED. Second, the metal outflow can continue to accelerate only if the outflow does not become over-ionized after decoupling from the H gas. This requires $\log U\la1$ for the metal outflow, for the AGN SED, so that at least $\sim10^{-3}$ of C and N can absorb in \CIV\ and \NV, respectively (e.g., \citealt{hamann97}). Consolidation of the two conditions requires a radiation-filtering shield that is further discussed below.

The increase in $U$ of the metal outflow is caused by two processes.
\begin{enumerate}
\item Once the metal outflow leaves its origin, the total number density of free particles drops considerably. This leads to a large increase of $U$. The density drop is of the order of magnitude of $\sim Z/20$, where the factor of 20 represents the mean atomic weight of metals. Note that due to the lower number density the recombination rate is also lower (the loss of H means a loss of $\sim10^2$ of the free electrons). The electron density drop leads to an increase in $U$ by $\Delta U\sim2$~dex.
\item The acceleration of the outflow gas causes an additional decrease of the gas number density $n_{\rm gas}$. Since the thermal velocity of the outflow gas ($b\sim10$~\kms) is much smaller than the outflow velocity ($v\sim10^3$~\kms), the expansion of the outflow perpendicular to its acceleration can be neglected, and the continuity equation yields $n_{\rm gas}\propto v^{-1}$.  Taking $v_{\rm term}$ to be $\sim 30\times10^3$~\kms, defining the origin of the high metal/H outflow by $w$ for which the metals run away ($w\sim 10^3$~\kms; Fig.~\ref{fig:dyn_fric})\footnote{We use $w$ rather than the typical velocity $v_{\rm typ}$ of the gas (e.g., a BLR cloud), since $v_{\rm typ}$ is probably due to the Keplerian motion of the gas, and is perpendicular to the $a_{\rm rad}$ which is probably radial.}, and noting that the distance travelled by the gas until $v_{\rm term}$ is reached [$r\sim\frac{1}{2}v_{\rm term}^2/a_{\rm rad}\sim(30\times10^8)^2/10^6\sim10^{13}$~cm; the value of $a_{\rm rad}$ is adopted from Fig.~\ref{fig:dyn_fric}] is probably much smaller than the distance from the ionizing source ($r\ga10^{15}$~cm) i.e., the ionizing flux is constant throughout the outflow, yields a change in $U\propto n_{\rm gas}^{-1}\propto v$ of $\Delta U\sim\log v_{\rm term}/w=1.5$~dex.
\end{enumerate}   
A combined effect of the two processes is an increase of $U$ by $\Delta U\sim 2+1.5=3.5$~dex. 

A shield, which filters the ionizing photons (i.e., lowers $U$) without attenuating significantly the UV flux that is absorbed by the metals, is needed to produce a metal runaway [e.g., the shield should produce $\log f_\nu({\rm UV})/n_{\rm H}\ga-10$, $T=10^4$~K and an \emph{effective} $\log U\la0$]; and to prevent the resulting high metal/H outflow to become over-ionized, and as a consequence to stall. The shield should filter the harder radiation that can over-ionize the metals (e.g., \citealt{murrayetal95}). Note that \citet{chenet03b} and \citet{everett05} also invoke a shield while modelling a radiation-driven wind. They find that a shielded outflow is accelerated more efficiently.

There are two possible physical scenarios which can produce a high metal/H outflow without invoking an additional shielding component. 
\begin{enumerate}
\item If the outflow is launched from the BLR, then the BLR gas might provide the needed filtering. The outflow might originate on the backside of the BLR cloud, behind most of the He$^+$ ionization front for C$^{3+}$ and N$^{4+}$ outflows, and part of the H$^0$ ionization front for a Si$^{3+}$ outflow. The filtering is a result of the bound-free absorption edge, for example by He$^+$  at 4~Ryd. The filter must allow production of N$^{4+}$, C$^{3+}$ and Si$^{3+}$, and thus it should have $\tau<1$ for photon energies $E>5.7$, 3.5 and 2.5~Ryd, respectively. The filter also has to prevent further ionization of these species i.e., $\tau\gg1$ for $E>7.2$, 4.7 and 3.3~Ryd. Thus, a strong He$^+$ edge, which is typically formed inside the BLR clouds, will form the right condition for the formation and acceleration of the C$^{3+}$ ion.
\item If the outflow is launched from an atmosphere of an accretion disc (AD), then the accelerating radiation (UV) is found much closer to the gas, while the over-ionizing radiation (EUV and above) is produced at smaller disc radii (e.g., \citealt{sands73}). Both radiation sources are anisotropic. Thus, the gas can be illuminated by a sufficiently large $f_\nu$ to produce an ion runaway, while maintaining a small enough ionizing $U$ that does not cause the outflow to be over-ionized. For example, for a local blackbody with temperature $T_{\rm disc}$, one requires that $n_\gamma$ that can over-ionize the ion (e.g., ionize C$^{3+}$ to C$^{4+}$) is smaller than the specie number density i.e., $n_\gamma\la10^{-4}n_p$ for a $Z_{\sun}$ gas. We use $n_\gamma\approx \nu f_\nu(\lambda_{\rm th})/h\nu c$, where $\lambda_{\rm th}$ is the threshold wavelength required to over-ionize the specie. Inserting in Eq.~\ref{eq:runaway} for $\log T =4$ yields $\log f_\nu(\lambda_{\rm ion})/f_\nu(\lambda_{\rm th})>9.1$ i.e., $(\nu_{\rm ion}/\nu_{\rm th})^3\exp[-h(\nu_{\rm ion}-\nu_{\rm th})/k T_{\rm disc}]>10^{9.1}$, where $\nu_i\equiv c/\lambda_i$ and the blackbody radiation is well approximated by the Wien law. This condition holds for $T_{\rm disc}\la35,000$, 25,000 and 15,000~K for \NV, \CIV\ and \SiIV, respectively. At these temperatures the local blackbody emission may be strong enough to produce and accelerate the ions, but too weak to destroy them.
\end{enumerate}
The metal outflow over-ionization is readily prevented in the BLR scenario. The outflow accelerates along an approximately straight trajectory defined by the illuminating source and the BLR cloud, and the cloud continues to filter the radiation reaching the accelerating outflow. If the metal outflow is launched from an AD, it can be accelerated only along a trajectory that is approximately parallel to the AD surface. The contribution of the AD inner parts to the total radiation illuminating the outflow increases as the outflow reaches larger altitudes above the disc surface. The radiation emitted by an AD becomes harder with a decreasing  disc radius \citep{sands73}. At a large enough altitude, the outflow is exposed to a hard enough total radiation that can over-ionize the gas, and the outflow stalls.

\subsection{Possible caveats to the metal decoupling scenario}\label{sec:dis_cav}
There are three possible caveats to the proposed metal decoupling scenario that should be considered.
\begin{enumerate}
\item If the gas at the outflow origin is embedded in closed loops of magnetic field that wrap the gas, then the ions will move along the field lines unable to run away from the gas. However, if the magnetic field lines extend away from the gas, then the metal ions might be accelerated along them and run away from the gas, possibly producing an observable effect.
\item The dynamical time for an ion runaway is $\sim\int_0^{v_{\rm th,e}}dw/[a_{\rm rad}-a_{\rm fric}(w)]$ which is of the order of 5~min for the case presented in Fig.~\ref{fig:dyn_fric} [$\log f_\nu(\lambda_{\rm ion})=-4.2$, $\log n_{\rm H}=6$ and $\log T=4$]. If there are turbulent processes that mix the gas on time scales smaller than this, then the ion acceleration might be stalled, preventing the runaway. However, it is not clear what mechanism can produce nearly supersonic turbulence  on the very small scales
of $10^{10}$~cm potentially involved with the acceleration.
\item The effect of a bulk motion of the accelerated ions through the H gas is considered below. A bulk acceleration of ions should produce a current density that acts as an additional deceleration force.\footnote{A current density produces a magnetic field. An interaction, between this magnetic field and charged particles that are coupled to the H gas, acts as a decelerating force on the bulk ions that produce the current density.} However, if the accelerating ions are shielded by free electrons of the gas, a current density is not produced. The length scale of acceleration of a given metal ion can be approximated by $h_{\rm acc} \approx 0.5v_{\rm th,ion}^2a_{\rm rad}^{-1}$. There is an electric shielding of the ion, if $h_{\rm acc}$ is much larger than the Debye shielding distance, $h_{\rm D}=\sqrt{k T/4\pi n_{\rm H} e^2}$, because then the ionized fluid (i.e., plasma) can effectively react to the ion motion. The condition of effective shielding is then $1\ll h_{\rm acc}/h_{\rm D}= 2.3\times10^{-5} f_{12}^{-1}f_\nu^{-1}(\lambda_{\rm ion})\sqrt{T n_{\rm H}}$ that can be approximated by $\log f_\nu-0.5\log T -0.5\log n_{\rm H}<-4$. This condition combined with the condition for a runaway (Eq.~\ref{eq:runaway}) yields $\log n_{\rm H}-3\log T<4$, which holds for a large parameter space ($\log n_{\rm H}\la12$ and $\log T\ga3$). The shielding is effective while the ion velocity relative to the fluid is $w\la v_{\rm th,e}$. For larger relative velocities ($w>v_{\rm th,e}$) the bulk ions need to ``carry'' the shielding electrons. This results in a lower Coulomb frictional force at a given $w$, since the effective charge of an ion and electrons it carries is smaller than $Z_{\rm ion}$ (see Eq.~\ref{eq:fric}).
\end{enumerate}

\section{Conclusions}
We analyse the conditions required for a radiation-pressure-driven pure metal outflow, and possible observational evidence for its existence. We find the following:
\begin{enumerate}
\item A metal ion embedded in gas subject to radiation pressure will run away when $\log f_\nu(\lambda_{\rm ion})-\log\nh+ \log T>-6.2.$ (Eq.~\ref{eq:runaway}). For an average AGN SED this can be converted to $\log U\ga5$.
\item To avoid overionization the gas must be protected by a shield, most likely in the form of a strong He$^+$ edge at 4~Ryd.
\item Photoionization models indicate for solar metallicity an absolute minimum of $N(\mbox{H$^0$})=2.5N(\mbox{Si$^{3+}$})$, implying $\tau(\Lya)\geq1.8\tau(\mbox{\SiIV\ $\lambda$1394})$, regardless of $U$ and $\Sigma$. 
\item Thus, a comparison of $\tau(\SiIV\,\lambda\lambda1394,\,1403)$ and $\tau(\Lya) $, for different absorption geometries, can yield a direct constraint on the metals/H abundance ratio.
\item A search of the SDSS BALQ sample of \citet{sca09} yields a handful of possible candidates for an outflow with a metallicity above solar. However, no robust evidence (i.e., $Z\ga10Z_{\sun}$) is found for a metal runaway.
\item High quality UV observation of lower $z$ quasars, free of intervening \Lya\ absorption, can be used to obtain better constraints on the outflow metallicity.
\item An ultra fast outflow of metals is expected to be produced if a runaway takes place.  Since the metal ions are not expected to be collisionally coupled, different ions will likely have  different terminal velocities.
\end{enumerate}

The physical conditions in AGN are definitely more complex than the simplistic scenario outlined in this paper. Clearly, models which simulate more accurately the various microphysics involved, including the caveats mentioned above, are required in order to get some quantitative predictions on metal ions runaway in radiation pressure driven outflows in AGN. Such models can then be integrated into simulations of the large scale structure of AGN outflows.

\section*{Acknowledgments}
We acknowledge fruitful discussions with N.\ Arav and J.\ H.\ Krolik. We thank the anonymous referee for many valuable comments, and G.\ Ferland for making \cl\ publicly available. AB thanks J.\ Stern for the help retrieving and reducing the SDSS spectra. This research has made use of the Sloan Digital Sky Survey which is managed by the Astrophysical Research Consortium for the Participating Institutions; of the VizieR catalog access tool, CDS, Strasbourg, France; and of the TOPbase on-line atomic database, which is a part of the Opacity Project.

\bsp
\label{lastpage}

\begin{thebibliography}{99}
\bibitem[\protect\citeauthoryear{Allen et al.}{2011}]{allenetal11} Allen J.\ T., Hewett P.\ C., Maddox N., Richards G.\ T., Belokurov V., 2011, MNRAS, 410, 860
\bibitem[\protect\citeauthoryear{Arav}{1996}]{arav96} Arav N., 1996, ApJ, 465, 617
\bibitem[\protect\citeauthoryear{Arav \& Li}{1994}]{arali94} Arav N., Li Z.-Y., 1994, ApJ, 427, 700
\bibitem[\protect\citeauthoryear{Arav \& Begelman}{1994}]{arabeg94} Arav N., Begelman M.\ C., 1994, ApJ, 434, 479
\bibitem[\protect\citeauthoryear{Arav, Li \& Begelman}{Arav et al.}{1994}]{aravetal94} Arav N., Li Z.-Y., Begelman M.\ C., 1994, ApJ, 432, 62
\bibitem[\protect\citeauthoryear{Arav et al.}{1999}]{aravetal99} Arav N., Korista K.\ T., De Kool M., Junkkarinen V.\ T., Begelman M.\ C., 1999, ApJ, 516, 27
\bibitem[\protect\citeauthoryear{Arav et al.}{2001}]{aravetal01} Arav N.\ et al., 2001, ApJ, 561, 118
\bibitem[\protect\citeauthoryear{Arav et al.}{2007}]{aravetal07} Arav N.\ et al., 2007, ApJ, 658, 829
\bibitem[\protect\citeauthoryear{Baskin \& Laor}{2008}]{bl08} Baskin A., Laor A., 2008, ApJ, 682, 110
\bibitem[\protect\citeauthoryear{Becker et al.}{2000}]{beckeretal10} Becker R.\ H., White R.\ L., Gregg M.\ D., Brotherton M.\ S., Laurent-Muehleisen S. A., Arav N., 2000, ApJ, 538, 72
\bibitem[\protect\citeauthoryear{Brandt, Mathur \& Elvis}{Brandt et al.}{1997}]{brandtetal97} Brandt W.\ N., Mathur S., Elvis M., 1997, MNRAS, 285, L25
\bibitem[\protect\citeauthoryear{Brandt, Laor \& Wills}{Brandt et al.}{2000}]{brandtetal00} Brandt W.\ N., Laor A., Wills B.\ J., 2000, ApJ, 528, 637
\bibitem[\protect\citeauthoryear{Chelouche \& Netzer}{2001}]{chenet01} Chelouche D., Netzer H., 2001, MNRAS, 326, 916
\bibitem[\protect\citeauthoryear{Chelouche \& Netzer}{2003a}]{chenet03a} Chelouche D., Netzer H., 2003a, MNRAS, 344, 223
\bibitem[\protect\citeauthoryear{Chelouche \& Netzer}{2003b}]{chenet03b} Chelouche D., Netzer H., 2003b, MNRAS, 344, 233
\bibitem[\protect\citeauthoryear{Cottis et al.}{2010}]{cottisetal10} Cottis C.\ E., Goad M.\ R., Knigge C., Scaringi S., 2010, MNRAS, 406, 2094
\bibitem[\protect\citeauthoryear{Crenshaw, Kraemer \& George}{Crenshaw et al.}{2003}]{crenshawetal03} Crenshaw D.\ M., Kraemer S.\ B., George I.\ M., 2003, ARA\&A, 41, 117
\bibitem[\protect\citeauthoryear{Cresci et al.}{2010}]{crescietal10} Cresci G., Mannucci F., Maiolino R., Marconi A., Gnerucci A., Magrini A., 2010, Nat, 467, 811
\bibitem[\protect\citeauthoryear{de Kool \& Begelman}{1995}]{dekbeg95} de Kool M., Begelman M.\ C., 1995, ApJ, 455, 448
\bibitem[\protect\citeauthoryear{Di Matteo, Springel \& Hernquist}{Di Matteo et al.}{2005}]{dimatteoetal05} Di Matteo T., Springel V., Hernquist L., 2005, Nat, 433, 604
\bibitem[\protect\citeauthoryear{Diamond-Stanic et al.}{2009}]{diast09} Diamond-Stanic A.\ M.\ et al., 2009, ApJ, 699, 782
\bibitem[\protect\citeauthoryear{Dietrich et al.}{2003}]{dietrichetal03} Dietrich M., Hamann F.,  Shields J.\ C., Constantin A., Heidt J., Jäger K., Vestergaard M., Wagner S.\ J., 2003, ApJ, 589, 722
\bibitem[\protect\citeauthoryear{Elvis et al.}{1994}]{elvisetal94} Elvis M., Wilkes B.\ J., McDowell J.\ C., Green R.\ F., Bechtold J., Willner S.\ P., Oey M.\ S., Polomski E., Cutri R., 1994, ApJS, 95, 1E
\bibitem[\protect\citeauthoryear{Everett}{2005}]{everett05} Everett J.\ E., 2005, ApJ, 631, 689
\bibitem[\protect\citeauthoryear{Fabian, Celotti \& Erlund}{Fabian et al.}{2006}]{fabianetal06} Fabian A.\ C., Celotti A., Erlund M.\ C., 2006, MNRAS, 373, L16
\bibitem[\protect\citeauthoryear{Ferland}{1999}]{ferland99} Ferland G., 1999, in Ferland G., Baldwin J., eds, ASP Conf.\ Ser.\ Vol.\ 162, Quasars and Cosmology. Astron.\ Soc.\ Pac., San Francisco, p.\ 147
\bibitem[\protect\citeauthoryear{Ferland et al.}{1998}]{fer98} Ferland G.\ J., Korista K.\ T., Verner D.\ A., Ferguson J.\ W., Kingdon J.\ B., Verner E.\ M., 1998, PASP, 110, 761
\bibitem[\protect\citeauthoryear{Gallazzi et al.}{2005}]{gallazzietal05} Gallazzi A., Charlot S., Brinchmann J., White S.\ D.\ M., Tremonti C.\ A., 2005, MNRAS, 362, 41
\bibitem[\protect\citeauthoryear{Gibson et al.}{2009}]{gibsonetal09} Gibson R.\ R.\ et al., 2009, ApJ, 692, 758
\bibitem[\protect\citeauthoryear{Hamann}{1997}]{hamann97} Hamann F., 1997, ApJS, 109, 279
\bibitem[\protect\citeauthoryear{Hamann}{1998}]{hamann98} Hamann F., 1998, ApJ, 500, 798
\bibitem[\protect\citeauthoryear{Hamann \& Ferland}{1999}]{hamfer99} Hamann F., Ferland G., 1999, ARA\&A, 37, 487
\bibitem[\protect\citeauthoryear{Hamann et al.}{1997}]{hamannetal97} Hamann F., Barlow T.\ A., Junkkarinen V., Burbidge E.\ M., 1997, ApJ, 478, 80
\bibitem[\protect\citeauthoryear{Hamann et al.}{2002}]{hamannetal02} Hamann F., Korista K.\ T., Ferland G.\ J., Warner C., Baldwin J., 2002, ApJ, 564, 592
\bibitem[\protect\citeauthoryear{Hamann et al.}{2011}]{hamannetal11} Hamann F., Kanekar N., Prochaska J.\ X., Murphy M.\ T., Ellison S., Malec A.\ L., Milutinovic N., Ubachs W., 2011, MNRAS, 410, 1957
\bibitem[\protect\citeauthoryear{Jannuzi et al.}{1996}]{jannuzietal96} Jannuzi B.\ T.\ et al., 1996, ApJ, 470, L11
\bibitem[\protect\citeauthoryear{Jones et al.}{2010}]{jonesetal10} Jones T., Ellis R., Jullo E., Richard J., 2010, ApJ, 725, L176
\bibitem[\protect\citeauthoryear{Junkkarinen, Burbidge \& Smith}{Junkkarinen et al.}{1983}]{junkkaetal83} Junkkarinen V.\ T., Burbidge E.\ M., Smith H.\ E., 1983, ApJ, 265, 51
\bibitem[\protect\citeauthoryear{Knigge et al.}{2008}]{kniggeetal08} Knigge C., Scaringi S., Goad M.\ R., Cottis C.\ E., 2008, MNRAS, 386, 1426
\bibitem[\protect\citeauthoryear{K\"{o}nigl \& Kartje}{1994}]{konkar94} K\"{o}nigl A., Kartje J.\ F., 1994, ApJ, 434, 446
\bibitem[\protect\citeauthoryear{Korista et al.}{1996}]{koristaetal96} Korista K., Hamann F., Ferguson J., Ferland G., 1996, ApJ, 461, 641
\bibitem[\protect\citeauthoryear{Kulsrud}{2005}]{kul05} Kulsrud R.\ M., 2005, Plasma Physics for Astrophysics. Princeton Univ.\ Press, Princeton, NJ
\bibitem[\protect\citeauthoryear{Kurosawa \& Proga}{2008}]{kurpro08} Kurosawa R., Proga D., 2008, ApJ, 674, 97
\bibitem[\protect\citeauthoryear{Kwan}{1990}]{kwan90} Kwan J., 1990, ApJ, 353, 123
\bibitem[\protect\citeauthoryear{Lamers \& Cassinelli}{1999}]{lamcas99} Lamers H.\ J.\ G.\ L.\ M., Cassinelli J.\ P., 1999, Introduction to Stellar Winds. Cambridge Univ.\ Press, Cambridge, UK
\bibitem[\protect\citeauthoryear{Laor \& Brandt}{2002}]{laorbrandt02} Laor A., Brandt W.\ N., 2002, ApJ, 569, 641
\bibitem[\protect\citeauthoryear{Laor et al.}{1997}]{laoretal97} Laor A., Fiore F., Elvis M., Wilkes B.\ J., McDowell J.\ C., 1997, ApJ, 477, 93
\bibitem[\protect\citeauthoryear{Molina et al.}{2009}]{molinaetal09} Molina M.\ et al., 2009, MNRAS, 399, 1293
\bibitem[\protect\citeauthoryear{Moll et al.}{2007}]{molletal07} Moll R.\ et al., 2007, A\&A, 463, 513
\bibitem[\protect\citeauthoryear{Morton}{1991}]{mor91} Morton D.\ C., 1991, ApJS, 77, 119
\bibitem[\protect\citeauthoryear{Murray et al.}{1995}]{murrayetal95} Murray N., Chiang J., Grossman S.\ A., Voit G.\ M., 1995, ApJ, 451, 498 
\bibitem[\protect\citeauthoryear{Owocki \& Puls}{2002}]{owo02} Owocki S.\ P., Puls J., 2002, ApJ, 568, 965
\bibitem[\protect\citeauthoryear{Proga}{2000}]{proga00} Proga D., 2000, ApJ, 538, 684
\bibitem[\protect\citeauthoryear{Proga}{2003}]{proga03} Proga D., 2003, ApJ, 585, 406
\bibitem[\protect\citeauthoryear{Proga, Stone \& Drew}{Proga et al.}{1998}]{progaetal98} Proga D., Stone J.\ M., Drew J.\ E., 1998, MNRAS, 295, 595
\bibitem[\protect\citeauthoryear{Proga, Stone \& Drew}{Proga et al.}{1999}]{progaetal99} Proga D., Stone J.\ M., Drew J.\ E., 1999, MNRAS, 310, 476
\bibitem[\protect\citeauthoryear{Proga, Stone \& Kallman}{Proga et al.}{2000}]{progaetal00} Proga D., Stone J.\ M., Kallman T.\ R., 2000, ApJ, 543, 686
\bibitem[\protect\citeauthoryear{Reeves \& Turner}{2000}]{reetur00} Reeves J.\ N., Turner M.\ J.\ L., 2000, MNRAS, 316, 234
\bibitem[\protect\citeauthoryear{Reichard et al.}{2003}]{reichardetal03} Reichard T.\ A.\ et al., 2003, AJ, 126, 2594
\bibitem[\protect\citeauthoryear{Rodr{\a' i}guez Hidalgo et al.}{2007}]{rhidalgoetal07} Rodr{\a' i}guez Hidalgo P., Hamann F., Nestor D., Shields J., 2007, in Ho L.\ C., Wang J.-M., eds, ASP Conf.\ Ser.\ Vol.\ 373, The Central Engine of Active Galactic Nuclei. Astron.\ Soc.\ Pac., San Francisco, p.\ 287
\bibitem[\protect\citeauthoryear{Rodr{\a' i}guez Hidalgo, Hamann \& Hall}{Rodr{\a' i}guez Hidalgo et al.}{2011}]{rhidalgoetal11} Rodr{\a' i}guez Hidalgo P., Hamann F., Hall P., 2011, MNRAS, 411, 247
\bibitem[\protect\citeauthoryear{Rybicki \& Lightman}{2004}]{ryblih04} Rybicki G.\ B., Lightman A.\ P., 2004, Radiative Processes in Astrophysics. WILEY-VCH Verlag GmbH \& Co.\ KGaA, Weinheim, Germany
\bibitem[\protect\citeauthoryear{Scaringi et al.}{2009}]{sca09} Scaringi S., Cottis C.\ E., Knigge C., Goad M.\ R., 2009, MNRAS, 399, 2231
\bibitem[\protect\citeauthoryear{Schneider et al.}{2010}]{sch10} Schneider D.\ P.\ et al., 2010, AJ, 139, 2360
\bibitem[\protect\citeauthoryear{Shakura \& Sunyaev}{1973}]{sands73} Shakura N.\ I., Sunyaev R.\ A., 1973, A\&A, 24, 337
\bibitem[\protect\citeauthoryear{Shen et al.}{2011}]{shenetal11} Shen Y.\ et al., 2011, ApJS, 194, 45
\bibitem[\protect\citeauthoryear{Spitzer}{1962}]{spitz62} Spitzer L., 1962, Physics of Fully Ionized Gases, 2nd edition. John Wiley \& Sons, Inc., USA
\bibitem[\protect\citeauthoryear{Springmann \& Pauldrach}{1992}]{sp92} Springmann U.\ W.\ E., Pauldrach A.\ W.\ A., 1992, A\&A, 262, 515
\bibitem[\protect\citeauthoryear{Stocke et al.}{1992}]{stockeetal92} Stocke J.\ T., Morris S.\ L., W R.\ J., Foltz C.\ B., 1992, ApJ, 396, 487
\bibitem[\protect\citeauthoryear{Telfer et al.}{2002}]{tel02} Telfer R.\ C., Zheng W., Kriss G.\ A., Davidsen A.\ F., 2002, ApJ, 565, 773
\bibitem[\protect\citeauthoryear{Verner et al.}{1996}]{verneretal96} Verner D.\ A., Ferland G.\ F., Korista K.\ T., Yakovlev D.\ G., 1996, ApJ, 465, 487
\bibitem[\protect\citeauthoryear{Vitello \& Shlosman}{1988}]{vitshl88} Vitello P.\ A.\ J., Shlosman I., 1988, ApJ, 327, 680
\bibitem[\protect\citeauthoryear{Wang et al.}{2009}]{wangetal09} Wang T., Zhou H., Yuan W., Lu H.~L., Dong X., Shan H., 2009, ApJ, 702, 851
\bibitem[\protect\citeauthoryear{Weymann et al.}{1991}]{weymannetal91} Weymann R.\ J., Morris S.\ L., Foltz C.\ B., Hewett P.\ C., 1991, ApJ, 373, 23
\bibitem[\protect\citeauthoryear{Wu et al.}{2010}]{wuetal10} Wu J., Charlton J.\ C., Misawa T.,  Eracleous M., Ganguly R., 2010, ApJ, 722, 997
\bibitem[\protect\citeauthoryear{York et al.}{2000}]{york00} York  D.\ G.\ et al. 2000, AJ, 120, 1579
\bibitem[\protect\citeauthoryear{Yuan et al.}{2011}]{yuanetal11} Yuan T.-T., Kewley L.\ J., Swinbank A.\ M., Richard J., Livermore R.\ C., 2011, ApJ, 732, L14
\bibitem[\protect\citeauthoryear{Zarritsky, Kennicutt \& Huchra}{Zaritsky et al.}{1994}]{zaritskyetal94} Zaritsky D., Kennicutt R.\ C.\ Jr., Huchra J.\ P., 1994, ApJ, 420, 87
\end{thebibliography}
\end{document}